\documentclass[]{article}

\pdfoutput=1 
\usepackage{arxiv}
\usepackage{booktabs}
\usepackage{siunitx}
\usepackage{tikz}
\usepackage[disable]{todonotes}
\usepackage{cite}
\usepackage{color}
\usepackage[normalem]{ulem}
\usepackage{bm}
\usetikzlibrary{arrows,decorations,positioning,shapes.geometric}
\usepackage{hyperref}

\usepackage[acronym]{glossaries}

\makenoidxglossaries 
\newacronym{abft}{ABFT}{\emph{Algorithm-Based Fault Tolerance}}
\newacronym{amr}{AMR}{\emph{Adaptive Mesh Refinement}}
\newacronym{api}{API}{\emph{Application Programming Interface}}
\newacronym{blcr}{BLCR}{\emph{Berkeley Lab Checkpoint/Restart}}
\newacronym{cds}{CDs}{\emph{Containment Domains}}
\newacronym{crc}{CRC}{\emph{Cyclic Redundancy Checks}}
\newacronym{cpu}{CPU}{\emph{Central Processing Unit}}
\newacronym{dce}{DCE}{\emph{Detectable and Correctable Error}}
\newacronym{dls}{DLS}{\emph{Dynamic Loop Self-scheduling}} 
\newacronym{dmr}{DMR}{\emph{Double modular redundancy}}
\newacronym{dls4lb}{DLS4LB}{\emph{Dynamic Loop Scheduling for Load Balancing}}
\newacronym{dram}{DRAM}{\emph{Dynamic Random-Access Memory}}
\newacronym{dsl}{DSL}{\emph{Domain Specific Languages}}
\newacronym{due}{DUE}{\emph{Detectable, but Uncorrectable Error}}
\newacronym{ecc}{ECC}{\emph{Error Correcting Codes}}
\newacronym{ecp}{ECP}{\emph{Exascale Computing Project}}
\newacronym{fem}{FEM}{\emph{Finite Element Method}}
\newacronym{fpga}{FPGA}{\emph{Field-Programmable Gate Arrays}}
\newacronym{fft}{FFT}{\emph{Fast Fourier Transforms}}
\newacronym{hpc}{HPC}{\emph{High Performance Computing}}
\newacronym{gpu}{GPU}{\emph{Graphics Processing Units}}
\newacronym{gvr}{GVR}{\emph{Global View Resilience}}
\newacronym{hbm}{HBM}{\emph{High-Bandwidth Memory}}
\newacronym{lflr}{LFLR}{\emph{Local-Failure Local-Recovery}}
\newacronym{mds}{MDS}{\emph{Meta Data Service}}
\newacronym{mpi}{MPI}{\emph{Message Passing Interface}}
\newacronym{nvm}{NVM}{\emph{Non-Volatile Memory}}
\newacronym{ode}{ODE}{\emph{ordinary differential equations}}
\newacronym{os}{OS}{\emph{Operating Systems}}
\newacronym{pcg}{PCG}{\emph{Preconditioned Conjugate Gradient}} 
\newacronym{pde}{PDE}{\emph{Partial Differential Equations}}
\newacronym{pfs}{PFS}{\emph{Parallel File System}}
\newacronym{pmpi}{PMPI}{\emph{MPI Profiling Interface}}
\newacronym{pvfs}{PVFS}{\emph{Parallel Virtual File System}}
\newacronym{qos}{QOS}{\emph{Quality of Service}}
\newacronym{rdlb}{rDLB}{\emph{robust Dynamic Load Balancing}}
\newacronym{tmr}{TMR}{\emph{Triple Modular Redundancy}}
\newacronym{ulfm}{ULFM}{\emph{User Level Failure Mitigation}}
\newacronym{sdc}{SDC}{\emph{Silent Data Corruption}}
\newacronym{ssd}{SSD}{\emph{Solid-State Drives}}

\newcommand{\todURi}[1]
{\todo[color=cyan!25,inline]{\footnotesize{\bf Uli:} #1}}
\newcommand{\todUR}[1]%
{\todo[color=cyan!25]{\footnotesize{\bf Uli:} #1} 
}
\newcommand{\todoLG}[1]%
{\todo[color=cyan!25]{\footnotesize{\bf Luc:} #1} 
}

\title{Resiliency in Numerical Algorithm Design for Extreme Scale Simulations}

\author{
Emmanuel Agullo \\
         Inria \\
         \And
Mirco Altenbernd \\
      Universit\"{a}t Stuttgart \\
      \And
Hartwig Anzt \\
        KIT -- Karlsruher Institut f\"{u}r Technologie \\
        \And
Leonardo Bautista-Gomez \\
         Barcelona Supercomputing Center  \\
         \And     
Tommaso Benacchio \\
        Politecnico di Milano \\
        \And
Luca Bonaventura \\
     Politecnico di Milano   \\
     \And 
Hans-Joachim Bungartz \\
             TU M\"{u}nchen  \\
             \And
Sanjay Chatterjee \\
       NVIDIA Corporation  \\
       \And
Florina M. Ciorba \\
        Universit\"{a}t Basel  \\
        \And
Nathan DeBardeleben \\
       Los Alamos National Laboratory \\
       \And
Daniel Drzisga \\
       TU M\"{u}nchen  \\
       \And
Sebastian Eibl \\
          Universit\"{a}t Erlangen-N\"{u}rnberg \\
          \And      
Christian Engelmann \\
          Oak Ridge National Laboratory  \\
          \And
Wilfried N. Gansterer \\
         University of Vienna  \\
         \And
Luc Giraud \\
    Inria   \\
    \And
Dominik G\"{o}ddeke \\
        Universit\"{a}t Stuttgart \\
        \And  
Marco Heisig\\
      Universit\"{a}t Erlangen-N\"{u}rnberg  \\
      \And
Fabienne J\'{e}z\'{e}quel \\
         Universit\'{e} Paris 2 -- Paris \\
         \And        
Nils Kohl \\
     Universit\"{a}t Erlangen-N\"{u}rnberg \\
     \And
Xiaoye Sherry Li \\
       Lawrence Berkeley National Laboratory \\
       \And
Romain Lion \\
       University of Bordeaux  \\
       \And
Miriam Mehl\\
       Universit\"{a}t Stuttgart  \\
       \And
Paul Mycek \\
     Cerfacs   \\
     \And
Michael Obersteiner \\
        TU M\"{u}nchen  \\
        \And
Enrique S. Quintana-Ort\'{i} \\
        Universitat Polit\`ecnica de Val\`encia \\
        \And   
Francesco Rizzi \\
          NexGen Analytics  \\
          \And
Ulrich R\"{u}de \\
       Universit\"{a}t Erlangen-N\"{u}rnberg  \\
       \And
Martin Schulz \\
       TU M\"{u}nchen  \\
       \And
Fred Fung \\
     Australian National University  \\
     \And
Robert Speck \\
       J\"{u}lich Supercomputing Centre  \\
       \And
Linda Stals \\
      Australian National University  \\
      \And
Keita Teranishi \\
      Sandia National Laboratories -- California  \\
      \And
Samuel Thibault \\
       University of Bordeaux  \\
       \And
Dominik Th\"{o}nnes \\
        Universit\"{a}t Erlangen-N\"{u}rnberg  \\
        \And
Andreas Wagner \\
        TU M\"{u}nchen  \\
        \And
Barbara Wohlmuth \\
        TU M\"{u}nchen  \\
}

\begin{document}

\maketitle
\clearpage

\pagestyle{headings}

\begin{abstract}

This work is based on the seminar titled ``Resiliency in Numerical Algorithm Design for Extreme Scale Simulations'' held March 1-6, 2020 at Schloss Dagstuhl, that was attended by all the authors.
Advanced supercomputing is characterized by very high computation speeds at the cost of involving an enormous amount of resources and costs.
A typical large-scale computation running for 48 hours on a system consuming
20 MW, as predicted for exascale systems, would
consume a million kWh, corresponding to about 100k Euro in energy cost
for executing $10^{23}$ floating-point operations.
It is clearly unacceptable to lose the whole computation if any of the several million parallel processes fails during the execution.
Moreover, if a single operation suffers from a bit-flip error, should the whole computation be declared invalid? 
What about the notion of reproducibility itself: 
should this core paradigm of science be revised and refined for results that are obtained by large scale simulation?

Naive versions of conventional resilience techniques will not scale to the exascale regime: with a main memory footprint of tens of Petabytes, synchronously writing checkpoint data all the way to background storage at frequent intervals will create intolerable overheads in runtime and energy consumption. Forecasts show that the mean time between failures could be lower than the time to recover from such a checkpoint, so that large calculations at scale might not make any progress if robust alternatives are not investigated.

More advanced resilience techniques must be devised. The key may lie in exploiting both advanced system features as well as specific application knowledge. Research will face two essential questions: (1) what are the reliability requirements for a particular computation and (2) how do we best design the algorithms and software to meet these requirements? While the analysis of use cases can help understand the particular reliability requirements, the construction of remedies is currently wide open. One avenue would be to refine and improve on system- or application-level checkpointing and rollback strategies in the case an error is detected. Developers might use fault notification interfaces and flexible runtime systems to respond to node failures in an application-dependent fashion. Novel numerical algorithms or more stochastic computational approaches may be required to meet accuracy requirements in the face of undetectable soft errors. These ideas constituted an essential topic of the seminar.

The goal of this Dagstuhl Seminar was to bring together a diverse group of scientists with expertise in exascale computing to discuss novel ways to make applications resilient against detected and undetected faults. In particular, participants explored the role that algorithms and applications play in the holistic approach needed to tackle this challenge. 
This article gathers a broad range of perspectives on the role of algorithms, applications, and systems in achieving resilience for extreme scale simulations. The ultimate goal is to spark novel ideas and encourage the development of concrete solutions for achieving such resilience holistically.

\end{abstract}

\newpage
\tableofcontents 
\newpage
\printnoidxglossaries
\printacronyms
\newpage


\section{Introduction}
\label{sec:introduction}

Numerical simulation is the third pillar in science discovery at the same level as theory and experiments.
To cope with the ever demanding computational resources needed by complex simulations, the computational power of high performance computing systems continues to increase by using an ever larger number of cores or by specialized processing. 
On the technological side, the continuous shrinking of transistor geometry and the increasing complexity of these devices affect 
their sensitivity to external effects and thus diminish their reliability.
A direct consequence is that \acrfull{hpc} applications are increasingly prone to errors.
Therefore the design of resilient systems and numerical algorithms that are 
able to exploit possible unstable \acrshort{hpc} platforms has became a major concern in the computational science community.
To tackle this critical challenge on the path to extreme scale computation an holistic and multidisciplinary approach is required that needs to involve researchers from various scientific communities ranging from the hardware/system community to applied mathematics for the design of novel numerical algorithms.
\nocite{heraultRobert:15}
In this article, we summarize and report on the outcomes
of a Dagstuhl seminar held March 1-6, 2020,\footnote{\url{https://www.dagstuhl.de/en/program/calendar/semhp/?semnr=20101}}
on the topic \emph{Resiliency in Numerical Algorithm Design for Extreme Scale Simulations}.
We should point out that, although error and resiliency was already quoted by J. von Neumann in his first draft report on EDVAC~\cite[P.1, Item 1.4]{EDVAC}, it became again a central concern for the HPC community in the late 2000' when the availability of the first exascale computers was envisioned for the forthcoming  decades. 
In particular, several workshops were organized in the IESP (International Exascale Software Project) and EESI (European Exascale Software Initiative) framework~\cite{cappelloetal2009}.

The  hardware/system resilience community has previously defined terminology related to how faults, errors, and failures occur on computing systems~\cite{AvizienisTaxonomy}.
In this article
our focus is less on the cause of an error (or the underlying fault), and more on how an error presents itself at the algorithmic level (or layer), impacting algorithms and applications.
We thus simplify the terminology often used in the hardware resilience and fault-tolerance community by not using terms like soft error or hard error, and generally do not concern ourselves with the reproducibility of an error (e.g., transient, intermittent or permanent).
This abstraction keeps the algorithmic techniques discussed herein general and applicable to a variety of fault models, current architectures, and hopefully of use in future technologies.

To this end, we broadly categorize errors presenting themselves to the algorithmic layer as either detected or undetected.  
Note that this categorization does not mean an error is undetect\underline{able} but rather that when it reached the algorithmic layer it was not detected by earlier layers (e.g., hardware, operating system or middleware/system software). 
This suggests the algorithmic layer has the opportunity to detect a previously undetected error and, if possible, to deploy mitigation methods to make the algorithm resilient; effectively transforming an undetected error at the algorithmic layer into a detected error.  
This in turn may result in a failure if the algorithm is unable to handle it.  
For example, an undetected data corruption which results in an application accessing an incorrect memory address may be detectable by the algorithm but it may not be possible for the algorithm to recover what the original memory address was and it may be forced to fail.  
If the algorithm could not detect the corruption before accessing the memory region, this would conventionally end in a failure (e.g., \texttt{SIGSEGV} issued by the operating system).

Many computing-intensive scientific applications that are dependent on \acrshort{hpc} performance upgrades can end up with disrupted schedules because of lack of resilience. A typical example is related to current efforts towards exascale numerical weather prediction~\cite{BenacchioEtAl2020, BenacchioEtAl2020_2}. On one side, regular upgrades in weather forecast models in operations at weather centres and their spatial resolution have gone hand in hand with expanding computational resources. 
On the other side, scientific and socioeconomic significance of forecasts crucially hinges on tight time-bound computing schedules and timely forecast dissemination, most notably for high-impact weather events. Current disk-checkpointing schedules still take up acceptable portions of forecast runtimes, but are hardly sustainable - indeed, they already saturate file systems bandwidth. In addition, many weather forecast codes feature preconditioned iterative solvers of linear systems with several hundred thousand unknowns, many thousand times per run. Such components represent vulnerable points in a context of increasingly frequent detected and undetected errors. Novel low-overhead solutions to enhance algorithmic fault-tolerance or provide higher-level system resilience are therefore in high demand in this and other fields where nonlinear dynamics is simulated.

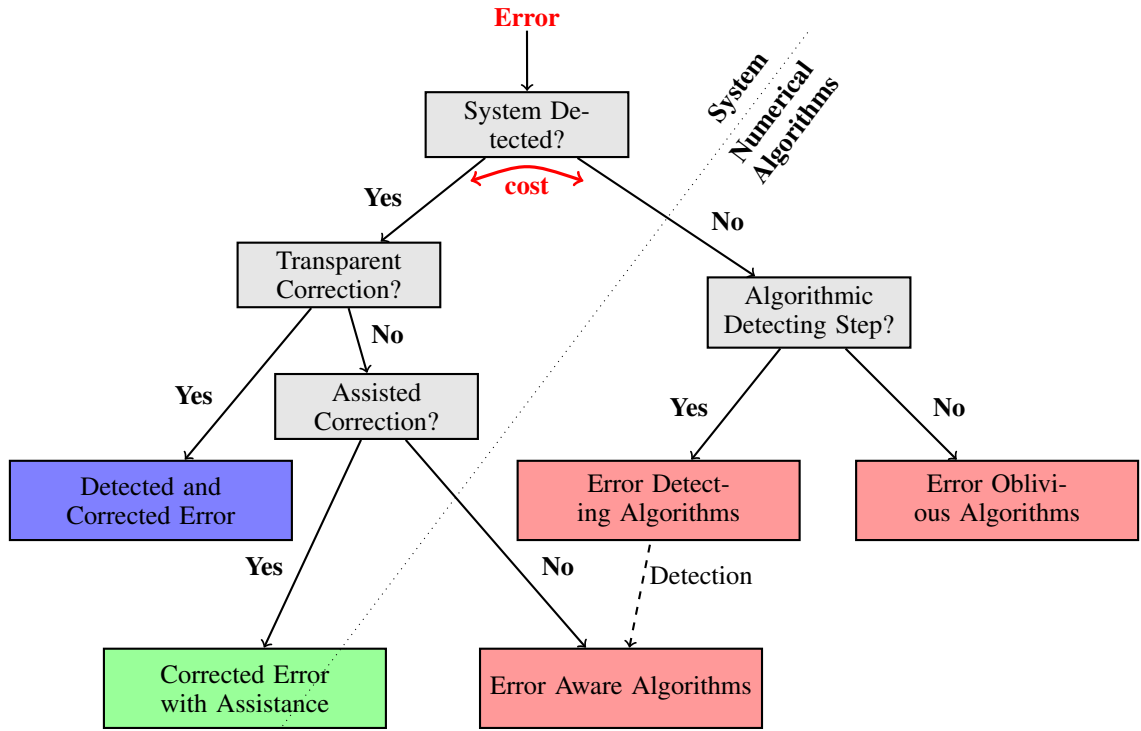
\begin{figure}
    \centering
      \begin{tikzpicture}
        [scale=2.5, allow upside down,
        box/.style={aspect=2,rectangle,draw,fill=black!10,thick,align=center,text width=7em},
        result/.style={rectangle,fill=black!10,thick,minimum size=3em,align=center,text width=10em},
        edge/.style={thick},
        ]
        \node (X) at (3,4) [box]                        {System Detected?};
        
        \node (Y) at (2,3.2) [box]                      {Transparent Correction?};
        \node (W) at (4.5,3) [box]                      {Algorithmic Detecting Step?};

        \node (Z) at (2.2,2.5) [box]                    {Assisted Correction?};

        \node (B) at (1,2) [box, result,fill=blue!50]        {Detected and Corrected Error};
           
        \node (F) at (3.7,2) [box, result,fill=red!40]       {Error Detecting Algorithms};
        \node (E) at (5.5,2) [box, result,fill=red!40]       {Error Oblivious Algorithms};

        \node (C) at (1.5,1) [box, result,fill=green!40]    {Corrected Error with Assistance};
        \node (D) at (3.5,1) [box, result,fill=red!40]       {Error Aware Algorithms};

        \draw [<->,red, very thick] (2.7,3.7) .. controls (3.0,3.8) .. (3.3,3.7) node [below,sloped,pos=0.5] {\textbf{cost}};
        
        \draw [dotted] (1.7,0.8) -- (4.5,4.5) node [above,sloped,pos=0.9] {\textbf{System}} node [below,sloped,pos=0.9,align=center] {\textbf{Numerical} \\ \textbf{Algorithms}};
        
        \draw [edge,->] (3,4.5) -- (X) node [above,pos=0.1,text=red,] {\textbf{Error}};
        
        \draw [edge,->] (X) -- (Y) node [above left,pos=0.7] {\textbf{Yes}};
        \draw [edge,->] (X) -- (W) node [above right,pos=0.7] {\textbf{No}};
        
        \draw [edge,->] (Y) -- (B) node [above left,pos=0.7] {\textbf{Yes}};
        \draw [edge,->] (Y) -- (Z) node [above right,pos=0.7] {\textbf{No}};
        \draw [edge,->] (W) -- (F) node [above left,pos=0.7] {\textbf{Yes}};
        \draw [edge,->] (W) -- (E) node [above right,pos=0.7] {\textbf{No}};

        \draw [edge,->] (Z) -- (C) node [above left,pos=0.7] {\textbf{Yes}};
        \draw [edge,->] (Z) -- (D) node [above right,pos=0.7] {\textbf{No}};
        \draw [edge,<-,dashed] (D) -- (F) node [above right,pos=0.5,align=center] {Detection};
    \end{tikzpicture}

    \caption{A classification of error handling.}
    \label{fig:error-handling-classification}
\end{figure}

In this article we take a different approach at the classification of errors in \acrshort{hpc} 
systems. 
In general, we try to divide errors in two main groups, those that are detected and corrected by the hardware/system (which is the focus of Section~\ref{sec:infrastructure}) and those that are detected and sometimes corrected by the numerical algorithms (Section~\ref{sec:algorithms}). 
However, the \acrshort{hpc} resilience ecosystem is not black and white, but it rather shows a wide palette of greys in between, with multiple fault tolerance tools implemented at the middleware level that are assisted by the applications/algorithms and vice-versa.
Figure~\ref{fig:error-handling-classification} shows this wide range of different error classifications depending on how much effort is needed at the application/algorithmic level in order to detect/correct the error.

The first category we observe in the leftmost leaf of the tree (blue color) is the case of errors that are both detected and transparently corrected by the hardware/middleware but without any intervention of the applications/algorithms. 
The clearest example would be a detectable and correctable error
in the memory generated by a single bit flip. 
These types of errors are transparently corrected by the system without any knowledge at the application/algorithmic level that such error mitigation occurred. 
Other examples could be process replication, system-level checkpointing, process migration, among many others (see Section~\ref{sec:soa-transparent-corrected}).  

The second category is the case of errors that are detected at the hardware/system level and are mitigated at the system/middleware level (not at the algorithmic level) but with assistance from the application/algorithm (green color). 
The most clear example of this is application-based checkpointing libraries, which handle all or most of the data transfers between the compute nodes and the 
\acrfull{pfs}
independently from the application, but it gets hints from it to know what datasets need to be checkpointed and when should the checkpoint happen. 
Other relevant examples are fault tolerant message passing programming models and resilient asynchronous tasks. 
We divide these sections in those approaches that require just a minor addition in the application code versus those that require a complete change in the programming paradigm (see Section~\ref{sec:soa-assistance-corrected}).

The other leaves of the tree (red color) correspond to those errors that cannot be corrected or mitigated at the hardware/system level and have to be mitigated by changing the algorithm or numerical methods to be able to tolerate those errors. 
We observe three different types of algorithms in this branch of the tree. 

The first type of algorithms focuses on the mitigation of errors that have been detected (first red leaf from left to right), 
we call them error-aware algorithms. 
Please note that these algorithms are not in charge of detecting the errors but only of mitigating them. 
Also, it is important to notice that these algorithms do not depend on how the error was actually detected;
it could be hardware/middleware detection as well as algorithmic detection, in the end the process of detection is irrelevant for the mitigation algorithm (see Section~\ref{sec:error_aware_algorithms}). 

The second type  of algorithms are those dedicated to the detection of errors that were not detected at the lower levels (fourth leaf from left to right).
A good example would be \acrfull{sdc} errors that pass invisibly 
through the hardware but then can be caught at the 
algorithmic level using some numerical techniques (e.g., checksum). 
These algorithms do not try to mitigate the error per se but only detect it.
Once the error has been detected, it can be passed to a error aware algorithm in order to attempt a correction/mitigation
(see Section~\ref{sec:error_detection}). 

Finally, there also exist algorithms that can operate, tolerate and absorb errors without ever being aware that the error ever occurred (last leaf to the right); we called these, 
error oblivious algorithms. 
These are somehow similar to the very first (blue) category, in that the errors are transparently corrected/absorbed,
see Section~\ref{sec:error_oblivious}.

In the following sections we discuss algorithmic and application approaches to address these two categories of errors and distinguish how the approaches vary or are similar.
Broadly speaking, the report is divided into two parts.
In  Section~\ref{sec:infrastructure} and Section~\ref{sec:algorithms} we discuss the state-of-the-art in the areas of infrastructure and algorithms, while in  Section~\ref{sec:future-directions} we propose possible areas of interest in future research.

\section{System infrastructure techniques for resilience}
\label{sec:infrastructure}

In this section we describe the state-of-the-art of hardware and system level error detection and mitigation. As previously mentioned, we divide these methods in two categories, the ones that mitigate the error in a completely transparent fashion, and those that require assistance from the algorithmic/application level. The following subsection, Section \ref{sec:soa-transparent-corrected}, concentrates on the methods falling in the first category. The second category is explored in Section \ref{sec:soa-assistance-corrected}.

\subsection{Detected and transparently corrected errors} \label{sec:soa-transparent-corrected}

A wide range of errors can be detected and immediately corrected by various layers in the system, i.e., these errors become masked or absorbed and higher level layers do not have to be involved. The detection/correction mechanisms have an extra cost in terms of storage, processing and energy consumption. 

\paragraph*{Hardware reliability}

At the hardware level several techniques exist to detect and correct errors. Most common examples are \acrfull{ecc}
to detect and correct single bit-errors, 
\acrfull{crc} error correction for network packets or RAID-1 (or higher) for I/O systems. A more comprehensive discussion of these features can be found in the report ``Towards Resilient EU HPC Systems: A Blueprint'' by Radojkovic et al.~\cite{EU_HPC}.

\paragraph*{Operating system reliability}

\acrfull{os} have certain capabilities to interact with architectural resilience features, such as ECC and machine check exceptions. OSs are mostly concerned with resource management and error notification. However, some advanced OS resilience solutions exist  such as Mini-ckpts~\cite{fiala16mini-ckpts}. It is a framework that enables application survival despite the occurrence of a fatal operating system failure or crash. It ensures that the critical data describing a process is preserved in persistent memory prior to the failure. Following the failure, the OS is rejuvenated via a warm reboot and the application continues execution effectively making the failure and restart transparent. The mini-ckpts rejuvenation and recovery process is measured to take \SIrange{3}{6}{\second} and has a failure-free overhead of \SIrange{3}{5}{\percent} for a number of key \acrshort{hpc} workloads.

\paragraph*{System-level checkpoint/restart}

\acrfull{blcr}~\cite{hargrove06berkeley} is a system-level checkpoint/restart solution that transparently saves and restores process state. In conjunction with a 
\acrfull{mpi}~\cite{mpi} implementation, it can transparently save and restore the process states of an entire MPI application. An extension of \acrshort{blcr}~\cite{varma06scalable,wang07job,wang10hybrid2} includes enhancements in support of scalable group communication for MPI membership management, reuse of network connections, transparent coordinated checkpoint scheduling, a job pause feature, and full/incremental checkpointing. The transparent mechanism for job pause allows live nodes to remain active and roll back to the last checkpoint, while failed nodes are dynamically replaced by spares before resuming from the last checkpoint. A minimal overhead of 5.6\% is reported in case migration takes place, while the regular checkpoint overhead remains unchanged. 

The hybrid checkpointing technique~\cite{gioiosa2005transparent} alternates between full and incremental checkpoints: At incremental checkpoints, only data changed since the last checkpoint is captured. This results in significantly reduced checkpoint sizes and overheads with only moderate increases in restart overhead. After accounting for cost and savings, the benefits due to incremental checkpoints are an order of magnitude larger than the overheads on restarts.

\paragraph*{Silent Data Corruption (SDC) protection}

FlipSphere~\cite{fiala16flipsphere} is a tunable, transparent 
\acrfull{sdc} detection and correction library for \acrshort{hpc} applications. It offers comprehensive SDC protection for application program memory using on-demand memory page integrity verification. Experimental benchmarks show that it can protect \SIrange{50}{80}{\percent} of program memory with time overheads of \SIrange{7}{55}{\percent}.

\paragraph*{Proactive fault tolerance using process or virtual machine migration}

Proactive fault tolerance~\cite{engelmann09proactive,wang12proactive,nagarajan07proactive} prevents compute node failures from impacting running applications by migrating parts of an application, i.e., tasks, processes, or virtual machines, away from nodes that are about to fail. Pre-fault indicators, such as a significant increase in temperature, can be used to avoid an imminent failure through anticipation and reconfiguration. As computation is migrated away, application failures are avoided, which is significantly more efficient than checkpoint/restart if the prediction is accurate enough. The proactive fault tolerance framework consists of process and virtual machine migration, scalable system monitoring and online/offline system health analysis. The process-level live migration supports continued execution of applications during much of process migration and is integrated into an MPI execution environment. Experiments indicate that \SIrange{1}{6.4}{\second} of prior warning are required to successfully trigger live process migration, while similar operating system virtualization mechanisms require \SIrange{13}{24}{\second}. This error oblivious approach complements checkpoint/restart by nearly cutting the number of checkpoints by half when 70\% of the faults are handled proactively.



\paragraph*{Resiliency using task-based runtime systems}
\label{task-based-runtime}

Task-based runtime systems have appealing intrinsic features for resiliency due to the fault isolation they provide by design as they have a view of the task flow and dynamically schedule task on computing units (often to minimize the time to solution or energy consumption). 
Once an error is detected and identified by the hardware or the algorithm, the runtime system can limit its propagation through the application by reasoning about the data dependencies among tasks~\cite{ocrhpec}.
For example, one can envision the scenario where an uncorrectable hardware error is detected triggering the runtime system to dynamically redistribute the tasks to the remaining resources available. 


Task-based runtime systems can also limit the size of the state needed to be saved to enable restarting computations, when an error is encountered~\cite{thibault:tel-01959127,lion2020,lion:hal-02296118}.
In classical checkpoint-restart mechanisms, the size of the checkpoint can become very large for large-scale applications, and managing it can take up a significant portion of the overall execution. 
A task-based runtime system simplifies the identification of points during the application execution when the state size is small, since only task boundaries need to be considered for saving the state.   
Further, identification of idempotent tasks can greatly help task-based runtimes to further reduce the overheads by completely avoiding data backups specific to those tasks. Recent works on on-node task parallel programming models suggest that a simple extension of the existing task-based programming framework enables efficient localized recovery  \cite{subasi2015nano,subasi2016runtime,paul2019hclib}.   

The checkpointing itself can also be achieved completely asynchronously~\cite{thibault:tel-01959127,lion2020,lion:hal-02296118}. The runtime allows tasks to read data being saved, and only blocks those tasks that attempt to overwrite data being saved. 
Since the runtime system knows which data will soon be overwritten by some tasks, it can prioritize the writing of the different pieces so as to have as little impact on the execution as possible. 
At the restarting point, the runtime also has all information to be able to achieve a completely local recovery. 
The replacement node can restart from the last valid checkpoint of the previously-failed node, while the surviving nodes can just replay the required data exchanges.

With the recent emergence of heterogeneous computing systems utilizing
\acrfull{gpu},
the task programming model is being used to offload computation from the 
\acrfull{cpu}
to the GPU. VOCL-FT~\cite{7832845} offers checkpoint/restart for computation offloaded to GPU using OpenCL~\cite{opencl}. 
It transparently intercepts the communication between the originating process and the local or remote GPU to automatically recover from \acrshort{ecc} errors experienced on the GPU during computation. 
Another preliminary prototype design extends this concept in the context of OpenMP~\cite{openmp} using a novel concept for \acrfull{qos} and a corresponding \acrfull{api}~\cite{engelmann19concepts}. 
While the programmer is specifying the resilience requirements for certain offloaded tasks, the underlying programming model runtime decides on how to meet them using a \acrshort{qos} contract, such as by employing task-based checkpoint-restart or redundancy.

\paragraph*{Resilience via complete redundancy}

The use of redundant MPI processes for error detection has been widely analyzed in the last decade~\cite{j137,thread-rep1,thread-rep2,thread-rep3}. Modular redundancy incurs high overhead, but offers excellent error detection accuracy and coverage with few to no false positive or false negatives.

Complete modular redundancy is typically too expensive for actual \acrshort{hpc} workloads.  However, it can make sense for certain subsystems such as parts of a \acrshort{pfs}.  The \acrfull{mds} of a networked \acrshort{pfs} is a critical single point of failure.  An interruption of service typically results in the failure of currently running applications utilizing its file system. A loss of state requires repairing the entire file system, which could take days on large-scale systems, and may cause permanent loss of data.  PFSs such as Lustre~\cite{lustre} often offer some type of active/standby fail-over mechanism for the MDS.  A solution~\cite{he09symmetric} for the MDS of the \acrlong{pvfs} offers symmetric active/active replication using virtual synchrony with an internal replication implementation.  In addition to providing high availability, this solution is taking advantage of the internal replication implementation by load balancing MDS read requests, improving performance over the non-replicated MDS.

\paragraph*{Resilience via partial redundancy}

Partial redundancy has been studied to decrease the overhead of complete redundancy~\cite{Elliott2012,ftxs12-rep,Subasi2015,Subasi2017}. 
Adaptive partial redundancy has also been proposed wherein a subset of processes is dynamically selected for replication~\cite{George2012}. 
Partial replication (using additional hardware) of selected MPI processes has been combined with \mbox{prediction-based} detection to achieve SDC protection levels comparable with those of full duplication~\cite{Berrocal16Exploring,Berrocal17Toward,FlipBack16}. A Selective Particle Replication 
approach for meshfree particle-based codes protects the data of the entire application (as opposed to a subset) by selectively duplicating \SIrange{1}{10}{\percent} of the computations within processes incurring a \SIrange{1}{10}{\percent} overhead~\cite{Cavelan:2019}.

\paragraph*{Resilience via complete and/or partial redundancy}

RedMPI~\cite{fiala12detection2} enables a transparent redundant execution of MPI applications. It sits between the MPI library and the MPI application, utilizing the \acrfull{pmpi} 
to intercept MPI calls from the application and to hide all redundancy-related mechanisms. A redundantly executed application runs with $r*m$ MPI processes, where $m$ is the number of MPI ranks visible to the application and $r$ is the replication degree. RedMPI supports partial replication, e.g., a degree of 2.5 instead of just 2 or 3, for tunable resilience.
It also supports a variety of message-based replication protocols with different consistency. Not counting in the need for additional resources for redundancy, results show that the most efficient consistency protocol can successfully protect \acrshort{hpc} applications even from high \acrshort{sdc} rates with runtime overheads from \SIrange{0}{30}{\percent}, compared to unprotected applications without redundancy. Partial and full redundancy can also be combined with checkpoint/restart~\cite{Elliott2012}. Non-linear trade-offs between different levels of redundancy can be observed when additionally using checkpoint/restart, since computation on non or less redundant resources is significantly less reliable than computation on fully or more redundant resources.



\paragraph*{Interplay between resilience and dynamic load balancing} 
\label{sec:runtime}

Scheduling of application jobs at the system level contributes to exploiting parallelism by placing and (dynamically) balancing the batch jobs on the local site resources. 
The jobs within a batch are already heterogeneous; yet, current batch schedulers rarely co-allocate, and most often only allocate, computing resources (while network and storage continue to be used as shared resources). 
Dynamic system-level parallelism can arise when certain nodes become unavailable (due to hard and permanent errors) or recover (following a repair operation). 
This can be exploited during execution by increasing opportunities for system-level co-scheduling in close proximity of jobs that exhibit different characteristics (e.g., co-scheduling a classical compute-intensive job in close proximity to a data-intensive job) and by dynamic resource reallocation to jobs that have lost resources due to failures or to waiting jobs in the queue.


\subsection{Detected errors mitigated with assistance} 
\label{sec:soa-assistance-corrected}

In this section we focus on correction methods that need assistance from the upper layers in order to achieve resilience and correctness. It is important to note that there are multiple methods that offer assisted fault tolerance but some of them involve a few additional lines of code while others require rewriting the whole applications using a specific programming model. Therefore, we will divide this section into subsections depending on the programming and/or redesign effort that is required. 

\subsubsection{Correction with incremental redesign} 
\label{sec:soa-correction-incremental}

As explained in Section \ref{sec:soa-transparent-corrected}, 
it is possible to perform system-level checkpointing without any feedback from the application or the algorithm or any upper layer.
The issue with system-level checkpointing is that the size (and therefore the time and energy cost) of checkpointing is much larger than what is really required to perform a restart of the application. 
Thus, application-level checkpointing is an attempt to minimize the size of checkpoints to the minimum required for the application to be able to restart.

\paragraph*{Performance modeling and optimization of checkpoint-restart methods}

Research on simulation tools assessing the performance of certain checkpoint-restart strategies is presented in various publications~\cite{levy2013using,
    engelmann2013toward,
    ashraf2018performance,
    di2014optimization}. 
Different theoretical approaches are used and tools are developed that either simulate a fictional software or wrap an actual application.

A lot of work has been done to examine and model the performance of multilevel checkpointing approaches~\cite{kohl2019scalable,zheng2012scalable,bautista2011fti,benoit2016optimal}.
Here, the parallel distribution of the snapshots as well as the target storage system are considered as objectives for performance optimization. 
Asynchronous techniques are considered, such as non-blocking checkpointing,
where a subset of processes are dedicated to manage the creation and reconstruction of snapshots~\cite{coti2006blocking,sato2012design}. 
As a measure to saving storage and speeding up I/O, data compression is another subject that is considered in the literature as, e.g., by Di and Cappello~\cite{di2016fast},
and in one of the case studies in Section~\ref{sec:error_aware_case_studies}.

Resilient checkpointing has been considered with the help of nonvolatile memory,
as for instance implemented in PapyrusKV~\cite{10.1145/3126908.3126943}, a resilient key-value blob-storage. 
Other resilient checkpointing techniques include the self-checkpoint technique~\cite{8170311}, which reduces common redundancies while writing checkpoints, or techniques reducing the amount of required memory through hierarchical checkpointing~\cite{5645453}, or differential checkpointing~\cite{DBLP:conf/ccgrid/KellerB19}.

\paragraph*{Message logging} \label{sec:logging}

Message logging is a mechanism to log communication messages in order to allow partial restart as for example examined by Cantwell et al.~\cite{cantwell:2019}.
While improving on basic checkpointing strategies, message logging-based approaches can themselves entail large overheads because of log sizes.
The checkpointing protocol developed by Ropars et al.~\cite{ropars:2013} does not require synchronization between replaying processes during recovery and limits the size of log messages. 
Other approaches combine task-level checkpointing and message logging with system-wide checkpointing~\cite{subasi:2018}.
This protocol features local message logging and only requires the restart of failing tasks.
It is also possible to combine message logging with local rollback and \acrfull{ulfm} (Section~\ref{sec:ULFM}) to improve log size~\cite{losada:2019}.



\paragraph*{Multilevel checkpointing libraries}\label{sec:multilevelCPR}

Current \acrshort{hpc} systems have deep storage hierarchies involving \acrlong{hbm}, 
\acrlong{dram}, \acrlong{nvm}, \acrlong{ssd}  and the \acrshort{pfs}, among others.
Multilevel Checkpointing libraries offer a way to leverage the different storage layers in the system through a simple interface. 
The objective is to abstract the storage hierarchy to the user,
so that one does not need to manually take care of where the data is stored or the multiple data movements required between storage levels. 
Each level of checkpointing provides a different trade-off between performance and resilience, 
where usually lower levels use close storage that offers higher performance but limited resilience,
and higher levels rely on stable storage (e.g., \acrshort{pfs}),
which is more resilient but slower. 
%
Mature examples of multilevel checkpoint libraries are SCR~\cite{SCR}, FTI~\cite{bautista2011fti}, CRAFT~\cite{shahzad2018craft} and VeloC~\cite{nicolae2019veloc}. Both SCR and FTI provide support via simple interfaces for storing application checkpoint data on multiple levels of storage, including RAM disk, burst buffers, and the parallel file system. Both SCR and FTI provide redundancy mechanisms to protect checkpoint data when it is located on unreliable storage and can asynchronously transfer checkpoint data to the parallel file system in the background while the application continues its execution. In addition, FTI also supports transparent GPU checkpointing. Finally, VeloC is a merge of the interfaces of both FTI and SRC. Note that some of these libraries offer the option for keeping multiple checkpoints so that the application can roll-back to different points in the past if necessary.

\paragraph*{Containment Domains}
\acrfull{cds} provide a programming construct to facilitate the preservation-restoration model, including nesting control constructs, and durable storage~\cite{container}. The following features are attractive for large-scale parallel applications. First, \acrshort{cds} respect the deep machine and application hierarchies expected in exascale systems. Second, \acrshort{cds} allow software to preserve and restore states selectively within the storage hierarchy to support local recovery. This enables preservation to exploit locality of storage, rather than requiring every process to recover from an error, and limits the scope of recovery to only the affected processors. Third, since \acrshort{cds} nest, they are composable. Errors can be completely encapsulated, or escalated to calling routines through a well-defined interface. We can easily implement hybrid algorithms that combine both preservation-restoration and data encoding.

Use cases include an implementation of a parallel resilient hierarchical matrix multiplication algorithm using a combination of ABFT (for error detection) and \acrshort{cds} (for error recovery)~\cite{Austin2015}. It was demonstrated that the overhead for error checking and data preservation using the \acrshort{cds} library is exceptionally small and encourages the use of frequent, fine-grained error checking when using algorithm based fault tolerance.

\paragraph*{Application versioning}

\acrfull{gvr}~\cite{DBLP:journals/ijhpca/ChienBDFFIRZHLR17} accommodates APIs to enable multiple versioning of global arrays for the single program, multiple data programming model.
The core idea is the fact that naive data redundancy approaches potentially store wrong applications states due to the large latency associated with error detection and notification. 
In addition to multiple versioning, GVR provides a signaling mechanism that triggers the correction of application states based on user-defined application error conditions. 
Use cases include an implementation of resilient Krylov subspace solvers~\cite{DBLP:conf/vecpar/ZhengCT14}.

\paragraph*{Mitigating performance penalties due to resilience via dynamic load balancing} \label{sec:DLS4LB}

Detected and corrected errors induce variation in the execution progress of applications when compared to error-free executions. 
This can manifest itself as load imbalance. 
Many application-level load balancing solutions have been proposed over the years
and  can help to address this problem.
We mention here a few available packages.

Available load balancing software includes
Zoltan~\cite{zoltan} 
that requires users to describe the workload across processes  as a
graph and offers an object oriented interface.
Further we mention \acrfull{dls4lb}~\cite{dls4l}, a recently developed library for MPI applications that contains a portfolio of self-scheduling based algorithms for load balancing.
StarPU~\cite{thibault:tel-01959127} proposes support for 
asynchronous load-balancing~\cite{lion2020} for task-based applications. 
The principle is to let the application submit only a part of its task graph, let some of it execute on the platform and observe the resulting computation balance. 
A new workload distribution can then be computed and the application is allowed to submit more of the task graph, whose execution can be observed as well.
OmpSs~\cite{OMPSS} is an effort to extend OpenMP in order to support asynchronous execution of tasks including a transparent interface for hardware accelerators such as 
\acrshort{gpu}s and \acrshort{fpga}s. 
OmpSs is built on top of the Mercurium compiler~\cite{mercurium} and the nanos++ runtime system~\cite{nanos}.

HCLib~\cite{yan2009hierarchical} is a task-based programming model that implements locality-aware runtime and work-stealing.
It offers a C and C++ interface and can be coupled with inter-process
communication models, such as MPI.
Charm++~\cite{CHARM} 
features an automatic hierarchical dynamic load balancing method that overcomes the scalability limitation of centralized load balancing as well as the poor performance of completely distributed systems. Such a technique can be triggered dynamically after a failure hits the system and the workload needs to be redistributed across workers.

\subsubsection{Correction with major redesign } 
\label{sec:ULFM}
The correction of some detected errors might have a strong impact of the algorithm that has to implement the mitigation. The mitigation design can be made more affordable if some components of the software stack have already some appealing features to handle such situations.
\paragraph*{Resilience support in the Message Passing Interface (MPI)}

Most MPI implementations by default are designed to terminate all processes when errors are detected.  However, this termination occurs irrespective of the scope of the error, requiring global shut-down and restart even for local errors in a single process.  This inherent scalability issue can be mitigated if MPI keeps all survived processes to continue and/or if restart overheads are reduced. The MPI community has proposed several recovery approaches, such as FA-MPI~\cite{hassani2014fampi} or MPI-ULFM~\cite{Bland:2012} to enable alternatives of global shut-down, as well as better error handling extensions, like MPI\_Reinit~\cite{laguna2016mpi}, to reduce overhead and impact of failures. 
Among these approaches, MPI-ULFM is the most advanced and well known. It provides a flexible low-level \acrshort{api} that allows application specific recovery via new error handling approaches and dynamic MPI communicator modification under process failures, although with significant complexities for the application developer using the new \acrshort{api}s. Several approaches have been proposed to mitigate this complexity  by creating another set of library \acrshort{api}s built atop of MPI-ULFM~\cite{cantwell:2019,gamell2014explore,Fenix,teranishi2014lflr,Shahzad2019craft}. However, as of now, in part due to its complexity when used on real-world applications and limited support in system software, MPI-ULFM as a whole has not been adopted in the MPI standard and hence is not readily usable for typical \acrshort{hpc} application programmers. Nevertheless, various aspects of \acrshort{ulfm} are in the process of standardization and will provide more mechanisms in MPI to build at least certain fault tolerant applications, starting with the upcoming MPI 4.0 standard.

\paragraph*{Resilience abstractions for data-parallel loops}

Data-parallel loops are widely encountered in $N$-body simulations, computational fluid dynamics, particle hydrodynamics, etc.  Optimizing the execution and performance of such loops has been the focus of a large body of work involving dynamic scheduling and load balancing. Maintaining the performance of applications with data-parallel loops running in computing environments prone to errors and failures is a major challenge. Most self-scheduling approaches do not consider fault-tolerance or depend on error and failure detection and react by rescheduling failed loop iterations (also referred to as tasks). A study of resilience in self-scheduling of data-parallel loops has been performed using SimGrid-based simulations of highly unpredictable execution conditions involving various problem sizes, system sizes, and application and systemic characteristics (namely, permanent node failures), that result in load imbalance~\cite{sukhija:2015}. Upon detecting a failed node, re-execution is employed to reschedule the loop iterations assigned to the failed node. 

A \acrfull{rdlb} approach has recently been proposed for the robust self-scheduling of independent tasks~\cite{Mohammed:2019}. The \acrshort{rdlb} approach proactively and selectively duplicates the execution of assigned chunks of loop iterations and does not depend on failure or perturbation detection. For exponentially distributed permanent node failures, a theoretical analysis shows that \acrshort{rdlb} is linearly scalable and its cost decreases quadratically with increasing system size. The reason is that increasing the number of processors increases the opportunities for selectively and proactively duplicating loop iterations to achieve resilience. \acrshort{rdlb} is integrated into a dynamic loop scheduling library (DLS4LB, see Section~\ref{sec:DLS4LB}) for MPI applications. \acrshort{rdlb} enables the tolerance of up to ($P-1$) process failures, where $P$ is the number of processes executing an application. For execution environments with performance-related fluctuations, \acrshort{rdlb} boosts the robustness of \acrfull{dls} techniques by a factor up to 30
and decreases application execution time up to 7 times compared to their counterparts without rDLB.

\paragraph*{Resilience extension for performance portable programming abstractions} \label{sec:soa-correction-major}

With the increasing diversity of the node architecture of \acrshort{hpc} systems, performance portability has become an important property to support a variety of computing platforms with the same source code while achieving a comparative performance to those programmed with the platform specific programming models. Today, Kokkos~\cite{edwards2014kokkos} and Raja~\cite{Raja2019, RajaGithub} accommodate modern C++ \acrshort{api}s to permit an abstraction of data allocation and parallel loop execution for a variety of runtime software and node architectures. This idea can be extended to express the redundancy of data and computation to achieve resilience while hiding the details of the data persistence and redundant computation. Recently, the resilient version of Kokkos was proposed for a natural API extension of Kokkos' data (memory space) and parallel loop (execution space) abstractions to (1) enable resilience with minimal code refactoring for the applications already written with Kokkos and (2) provide common interface to call any external resilience libraries such as VeloC~\cite{nicolae2019veloc}. The new software will be released in a special branch in \url{https://github.com/kokkos/kokkos}.

The resilience abstraction idea has also been applied to task parallel programming models such as Charm++~\cite{CHARM}, HClib~\cite{yan2009hierarchical}, HPX~\cite{HPX}, OmpSs~\cite{OMPSS} and StarPU~\cite{thibault:tel-01959127} to integrate a variety of resilient task program execution options such as replay, replication, algorithm-based fault tolerance and task-based checkpointing.
\todUR{Here could be refs included to load balancing, savimg some of the material from the deleted paragraphs abov}
Task-based programming models indeed have a very rich view over the structure of the application computation, and notably its data, and have a lot of control over the computation execution, without any need for intervention from the application. Replaying a failed task consists of issuing it again with the same input, discarding the previous erroneous output, and replicating a task consists of issuing it several times with different output buffers and comparing the result. Dynamic runtime systems can then seamlessly introduce replay and replication heuristics, such as trying to run different implementations and/or computation units, without the application having to be involved beyond optionally providing different implementations to be tried for the same task. 

The task graph view also allows for very optimized checkpointing~\cite{thibault:tel-01959127,lion2020,lion:hal-02296118}. In the task-based programming model, each checkpoint is a cut in the task graph, which can be expressed trivially within the task submission code, and only the data of the crossing edges need to be saved. Even better, the synchronization between the management of checkpoint data and application execution can be greatly relaxed. The transfer of the data to the checkpoint storage can indeed be started as soon as the data is produced within the task graph, and not only once all tasks before the checkpoint are complete. A checkpoint is then considered complete when all its pieces of data have been collected. It is possible that tasks occurring after the checkpoint may run to completion before the checkpoint itself is completed.
All in all, this allows for a lot more time for the data transfers to complete, and lessens the I/O bandwidth pressure.

\paragraph*{Software engineering approaches for resilience by design}

Resilience design patterns~\cite{hukerikar17rdp-12, hukerikar17pattern} offer an approach for improving resilience in extreme-scale \acrshort{hpc} systems. Frequently used in computer engineering, design patterns identify problems and provide generalized solutions through reusable templates. Reusable programming templates of these patterns can offer resilience portability across different \acrshort{hpc} system architectures and permit design space exploration and adaptation to different (performance, resilience, and power consumption) design trade-offs. An early prototype~\cite{ashraf18pattern-based} offers multi-resilience for detection, containment and mitigation of silent data corruption and MPI process failures.

\section{Numerical algorithms for resilience}
\label{sec:algorithms}

In this section, we focus on the handling of errors at the algorithmic level. We see three different classes of problems to tackle here: 
(i) detection of un-signaled errors (mostly bit flips and other instances of silent data corruption, Section~\ref{sec:error_detection}),
(ii) correction of errors that have been signaled but could not be corrected at the hardware or middleware layer (by error aware algorithms, Section~\ref{sec:error_aware_algorithms}), 
(iii) design of error oblivious algorithms that deliver the correct result even in the presence of (not too frequent) errors (Section~\ref{sec:error_oblivious}). 

In addition to correctness in the presence of errors, an important challenge in all our considerations is efficiency in terms of algorithm runtime. In this context, additional algorithmic components such as work stealing and asynchronous methods (where missing data are simply an extreme case of delay) have to be considered. We mention these methods when describing methods that can make use of such runtime optimizing measures.  

\subsection{Error detecting algorithms}  \label{sec:error_detection}


In this section, we focus on mechanisms to numerically detect errors that have not been detected by the underlying system or middleware. 
We have identified several techniques that allow us to (likely) notice the  occurrence of an error at several layers of numerical algorithms. 
Table~\ref{tab:detection_chart} gives an overview of some detection techniques and the algorithmic components or numerical methods where they are applicable. 
\begin{table}
    \centering
    \caption{Numerical error detection: Overview of error detection techniques and numerical ingredients and methods where they are applied. 
    Note that we mark a method as applicable only if it is or can be used in the respective algorithm itself, not only at lower level functionality, i.e., we do not mark checksums for multigrid as checksums are only used in the BLAS 2/3 kernels used as inner loops or in the GS/J/SOR smoothers.}
    \label{tab:detection_chart}
    \begin{tabular}{lcccccc}  
    & 
    \rotatebox[origin=l]{90}{exceptions} & \rotatebox[origin=l]{90}{checksum} & \rotatebox[origin=l]{90}{constraints} & \rotatebox[origin=l]{90}{tech error} & \rotatebox[origin=l]{90}{multi resolution} & \rotatebox[origin=l]{90}{redundancy} \\
    \midrule
    BLAS 2/3 & $\times$ & $\times$ &&&& $\times$  \\
    Direct Solvers & $\times$ & $\times$ &&&& $\times$  \\
    Krylov & $\times$ && $\times$ & $\times$ && $\times$  \\
    Multilevel / Multigrid & $\times$ &&& $\times$ & $\times$ & $\times$  \\
    Domain Decomposition & $\times$ &&&&& $\times$  \\
    GS/Jac/SOR & $\times$ &&& $\times$ && $\times$  \\
    Nonlinear Systems & $\times$ &&& $\times$ && $\times$  \\
    Time Stepping (ODEs) & $\times$ &&& $\times$ & $(\times)$ & $\times$  \\
    PDEs & $\times$ & $\times$ & $\times$ & $\times$ & $\times$ & $\times$  \\
    Quadrature & $\times$ && $\times$ & $\times$ & $\times$ & $\times$  \\
    \end{tabular}
\end{table}


\subsubsection{Exceptions}

Exceptions are a way a program signals that  something went wrong during execution.  We consider the case where exceptions are caused by data corruption that can, for example, lead to division by zero or out-of-range access. 
Most programming languages support a way of handling exceptions.
The algorithm programmer can register an exception handler that gets called whenever an exception occurs. If the error is recoverable, the exception handler will specify how best to continue afterwards. 
If the error is not recoverable, the program will be aborted.
Exceptions are a straight-forward way to detect certain types of errors and can be applied to all numerical algorithms. 
However, they obviously only see a small subset of all possible errors and it is not trivial to decide when to use exceptions handlers in the light of a trade-off between correctness, robustness and runtime efficiency.

\todURi{Here a description and/or references are missing how exceptions are used to deal with faults/errors/mitigation \textbf{DG: \cite{Engwer:2018:AHL} describes how to wrap ULFM into C++ exceptions, also in \cite{Altenbernd:2020:hpcse}}}


\subsubsection{Checksums} \label{sec:detect_checksum}

Checksums could be used at the hardware or middleware layer to detect errors, but here we will discuss checksums
as employed on the algorithmic layer where we have a more detailed knowledge about the existence of numerical or algorithmic invariants.
Checksum techniques have been used in various numerical algorithms. We list some examples below. 

\noindent
\textbf{BLAS 2/3:} 
Checksum encoding matrices, introduced by Huang and Abraham~\cite{Huang1984} requires (i) adding redundant data in some form (encoding), (ii) redesign of the algorithm to operate on the respective data structures (processing), and (iii) checking the encoded data for errors (detection). We ignore the recovery phase here and refer to Section~\ref{sec:error_aware_algorithms}. Checksums are used in FT-ScaLAPACK~\cite{Wu2014} for dense matrix operations such as MM, LU and QR factorization and more recently in hierarchical  matrix multiplication~\cite{Austin2015}. Wu et al.\ give a good survey of checksum deployment in dense linear algebra~\cite{PanGDBTLCC2016}.

\noindent
\textbf{Gauss-Seidel/Jacobi/SOR and multigrid:} 
In~\cite{Mishra:2003:Algorithm-Based}, checksums are used to detect errors in the Jacobi smoother, the restriction and interpolation operators of a multigrid method solving a two-dimensional Poisson equation. 

\noindent
\textbf{Krylov subspace methods:} 
Tao et al.\ propose a new checksum scheme using multiple checksum vectors for sparse matrix-vector multiplication, which is shown to be generally effective for several preconditioned Krylov iterative algorithms~\cite{TaoSKWLZKC2016}.
Also~\cite{Shantharam2011, agullo:hal-02495301} use checksums for protection within the conjugate gradient (CG) algorithm.
    
\noindent
\textbf{FFT:} 
Checksum can also be used in \acrfull{fft}s similarly as in matrix-vector multiplication. Liang et al.~\cite{LiangCTLWLOLSC2017} develop a new hierarchical checksum scheme by exploiting the special discrete Fourier transform matrix structure and employ special checksum vectors. Checksums are applicable to many important kernels such as matrix-matrix multiplication, but are costly.
In addition, it can be difficult to specify a suitable threshold for \lq equality\rq\ in the presence of round-off errors.
For many numerical calculations such as scalar products, checksums are not applicable at all.


\subsubsection{Constraints}

In some applications, constraints for different types of variables are known. Examples are positivity constraints, conservation laws for physical quantities or known bounds for internal numerical variables. 

\noindent
\textbf{Krylov subspace methods:}
Resilience was already of importance in the early days of digital computers. In the original \acrshort{pcg} paper~\cite{cg:52}, Hestenes and Stiefel noticed that the reciprocal value of $\alpha$ (the step length) is bounded above (repectively, below) by the reciprocal of the  smallest eigenvalues (respectively the inverse of the largest eigenvalue) of the matrix. The inequality involving the largest eigenvalue (for which in practice it may be cheaper to get an approximation) was used to equip \acrshort{pcg} with error detection capabilities in~\cite{agullo:hal-02495301}.

\noindent
\textbf{Partial differential equations:}
Checking for bounds can be associated with minimal or extremely high cost depending on whether extra information has to be computed (such as eigenvalues of matrices) or not. Reliability is, in general, an issue as only those errors leading to violation of these constraints can be detected. An example of the use of problem-informed constraints can be found in \cite{mycek:2017sisc}. In this work, the authors derive a priori bounds for the discrete solution of second-order elliptic \acrshort{pde}s in a domain decomposition setting. Specifically, they show that the bounds take into account the boundary conditions, are cheap to compute, general enough to apply to a wide variety of numerical methods such as finite elements or finite differences, and provide an effective way to handle faulty solutions synthetically generated.


\subsubsection{Technical error information} 
\label{sec:tec_err_info}

In many numerical large scale applications, the main computational task involves the approximate computation of integrals, algebraic systems, systems of  \acrshort{ode}s or \acrshort{pde}s. For all these problems, various types of error information such as residuals, differences between iterations, round-off error estimates and discretization error estimates can be used as indicators of errors either by their size or by monotonicity criteria. We give several examples from literature for different classes of numerical algorithms.

\noindent 
\textbf{Krylov subspace methods:}
Round-off error bounds can be used in Krylov subspace methods. They fit in the general framework of round-off error analysis~\cite{Higham96} and have been considered in the context of Krylov subspace methods in finite precision arithmetic~\cite{Strakos13,meurant2006lanczos}.

Vorst and Ye proposed a residual gap bound~\cite{Vorst2000} (bound for the norm of the residual gap between the true and the computed residuals) based on round-off error analysis that was later used as a criterion for actual error detection in~\cite{agullo:hal-02495301} when bit flips occur. The detection of errors in Krylov methods via violation of orthogonality is proposed in~\cite{CHen2013}.

\noindent
\textbf{Multigrid:}
Calhoun et.~al~\cite{Calhoun:2015:Towards} apply a residual/energy norm-based error detection for algebraic multigrid. They use two criteria: (i) the reduction of the residual norm as a weak criterion and (ii) the reduction of the quadratic form  \[ E(\bm{x}) = \langle A\bm{x}, \bm{x} \rangle - 2 \langle \bm{x}, \bm{b} \rangle, \] when solving the linear system $A \bm{x} = \bm{b}$ for symmetric positive matrices.

The quadratic for $E$ calculated at level $i$ during the down-pass of a V-cycle should be less than the energy calculated at level $i$ during the down-pass of the next V-cycle.

When using the full approximation scheme residual norm reductions can also be verified at each level in the hierarchy of a multigrid-cycle. 
The structure of the full approximation scheme additionally provides smart recovery techniques utilizing its lower resolution approximations~\cite{altenbernd2016fault}.

\noindent
\textbf{Time-stepping:} 
For iterative time-stepping with spectral deferred corrections, monitoring the residual of the iteration can be used to detect errors in the solution vectors~\cite{grout2017achieving}. In the context of parallel-in-time integration with parareal, consecutive iterates are considered in~\cite{nielsen2016fault} to detect errors in the solution vector. In~\cite{benson2015silent}, an auxiliary checking scheme in contrast to the original base scheme is used to detect and correct errors during implicit and explicit time-integration. Estimating the local truncation error with two different methods is used in~\cite{guhur2016lightweight} to implement a resilient, high-order Runge-Kutta method. This ``Hot Rod'' approach is then also used for error correction. 


\subsubsection{Multi-resolution}

Multi-resolution means that information is available at different resolution levels, in terms of spatial discretization (\acrshort{pde}), time discretization (\acrshort{ode} and \acrshort{pde}), order of discretization (PDE in space and time), matrix dimensions (numerical linear algebra, multigrid), frequencies, and so on. 
This leads to a certain redundancy -- not an artificially introduced, but an inherently available one. 
This redundancy can be used to detect discrepancies or anomalies and, hence, errors that could not be detected by the system. There are numerous examples for the mentioned problem classes, we outline one example in more detail here.

\noindent

\textbf{Sparse grids / Combination technique:} 
Sparse grids \cite{bungartz04sparseGrids} are one particular class of multi-resolution methods. 
There, via the use of hierarchical bases, certain structures often seen in $d$-dimensional data can be exploited to alleviate the curse of dimensionality, without a significant loss of accuracy. 
Sparse grids have been successfully used in a wide range of problem classes where spatial discretization plays a role, such as interpolation \cite{Jakeman2011}, quadrature \cite{GerstnerDimAdapt,gerstnerQuad, Bungartz2003}, solvers for PDEs \cite{heene2018exahd, harding2014robust}, or machine learning tasks \cite{Garcke2006, garcke2007dimension, PeherstorferSGDE} (e.g., classification, regression, clustering, or density estimation). 
One particular incarnation of sparse grid methods is the so-called combination technique \cite{griebel92CombiTechnique}. 
There, based on an extrapolation-style approach, a linear combination of a specific set of full, but very coarse-grid solutions is used to get a sparse fine-grid solution. 
The various coarse grid solutions can be obtained in a completely independent way, using (parallel) standard solvers. 
This opens the way to (1) a natural two-level parallelization and to (2) an easy and cheap detection of system undetected errors: Since we actually compute solutions for the same problem on different (i.e., differently discretized) grids anyway, we can use these to detect anomalies -- just by comparing the available solutions.  
And the detection leads immediately to a mitigation strategy (see Section~\ref{sec:error_aware_case_studies}), since we can easily exchange coarse grids in case of errors, just by changing the combination pattern \cite{obersteiner2017highly, ali-ft-gene-hpcs-2015, ali2016complex, heene2016massively, harding2015fault, parra16SDC}. 
Therefore, this is an example for a smart algorithm that is able to do both detection and mitigation.

Further examples are mentioned in Section~\ref{sec:tec_err_info} as multi-resolution typically comes with corresponding error estimates based on differences between solutions at different resolution levels: multigrid and parallel time stepping.


\subsubsection{Redundancy}

Redundancy is a strategy for error detection that can be applied to all of the numerical algorithms mentioned in Table~\ref{tab:detection_chart}.
It covers two approaches. 
In the first approach computational resources may be replicated twice or thrice. Such instances are called \acrshort{dmr} \cite{Weaver2001AFaultTolerant,Iyer2005recent} or \acrshort{tmr} \cite{vonNeumann56Proba,Scholzel2007Reduced}. 
In the second approach the computations are repeated twice or thrice on the same resource \cite{austin1999DIVA,Vijaykumar02transient-faultrecovery}. 
An advantage of this approach is the flexibility at the application level.
Note that the first approach costs more in space or resources, the second approach costs more in time.

The redundancy based error detection technique described in \cite{benoit2017replication} relies on in-depth analysis of application and platform dependent parameters (such as the number of processors and checkpointing time) to formalise the process of both resource and computation replication. 
It provides a closed-form formula for optimal period size, resource usage and overall efficiency.

Ainsworth et.\ al~\cite{ainsworth2017multigrid} use replication of fault-prone components as an error detection technique in a multigrid method. 
Also error detection in the time stepping methods from~\cite{benson2015silent} mentioned in Section~\ref{sec:tec_err_info} can be interpreted as redundancy based error detection.

The main disadvantage of replication is its cost in terms of performance,
although recomputing only some instructions instead of the whole application lowers the time redundancy overhead~\cite{Oh2002Error}.
However,  redundancy in some calculations should in particular be considered as a possible strategy for error detection as in modern supercomputers the cost of arithmetic operations tends to decrease compared to communication time.

\subsection{Error aware algorithms} \label{sec:error_aware_algorithms}

In this section, we look at error correction techniques within an application.
We assume that the application has been notified that part of the algorithm's data is corrupted or lost. 
In that context, mitigation or containment actions have to be undertaken at the algorithmic design level, where the appropriate actions depend on the data detection granularity and how the notification mechanism was activated. 
It is possible to design both lossy and lossless mitigation procedures that are tailored to the numerical algorithms under consideration.

In Section~\ref{sec:numalgos:erroraware:relatedwork} we give a brief literature overview of ideas that can be used to complement numerical mitigation or containment procedures. 
Then, in Section~\ref{sec:error_aware_case_studies} 
we offer a more detailed discussion of some recent successful attempts by presenting a few case studies in the context of the solution of \acrfull{pde}.


\subsubsection{Error aware algorithms for the solution of linear systems}\label{sec:numalgos:erroraware:relatedwork}


A wealth of literature already exists on various, mostly isolated ideas and approaches that have appeared over time. 
Checkpoint-restart methods are the most generic approaches towards resilience for a broad spectrum of applications, see Section~\ref{sec:multilevelCPR} for an introduction. 
We first describe a general mental model to design resilient numerical algorithms independent of actual machine specifications that lead to what is nowadays referred to as \acrfull{lflr} techniques. Then we move to \lq classical\rq\ algorithm-based fault tolerance, which originally was developed to detect and correct single bit flips on systolic architectures devoted to basic matrix computations, see Section~\ref{sec:detect_checksum}. Finally, we discuss a range of ideas and techniques not covered by the case studies below.

\paragraph*{Local-failure local-recovery}

As far back as a decade ago, an abstract framework was developed to separate algorithm design from unclear machine specifications, see also Section~\ref{sec:logging}. The idea of a selective reliability model as introduced by Hoemmen~\cite{Hoemmen2011, bridges2012faulttolerant} is machine-oblivious and highly suitable for algorithm design for machines with different levels of (memory) reliability. It has led to the concept of \acrfull{lflr}~\cite{teranishi2014lflr}. This model provides application developers with the ability to recover locally and continue application execution when a process is lost. In ~\cite{teranishi2014lflr}, Teranishi and Heroux have implemented this framework on top of MPI-ULFM (Section~\ref{sec:ULFM}) and analyzed its performance when a failure occurs during the solution of a linear system of equations.

\paragraph*{Original algorithm-based fault tolerance with checksums}

 The term \acrfull{abft} was originally coined in conjunction with protecting matrix operations with checksums to handle bit flips~\cite{huang:1984}, mostly assuming exact arithmetic calculation for detection and mitigation. 
 (See Section~\ref{sec:detect_checksum} for a more detailed discussion on checksums). 
 The main drawback of checksums is that only limited error patterns can be corrected and its robust practical implementation in finite precision arithmetic can be complicated to tune to account for round-off errors.
 A second drawback is that the checksum encoding, detection and recovery methods are specific to a particular calculation. 
 A new scheme needs to be designed and proved mathematically for each new operation. 
 A further drawback is to tolerate more errors, more encoded data is needed, which may be costly both in memory and in computing time.

\acrshort{abft} concepts have been extended to process failures for a wide range of matrix operations both for detection and mitigation purposes \cite{kim:1996,du:2012,bosilca:2009,chen:2008,jia:2013} and general communication patterns \cite{kabir:2016}. 
\acrshort{abft} has also recently been proposed for parallel stencil-based operations to accurately detect and correct silent data corruptions~\cite{Cavelan:2019a}. 
In these scenarios the general strategy is a combination of checkpointing and replication of checksums. 
In-memory checkpointing \cite{jia:2013} can be used to improve the performance.
The main advantage of these methods is their low overhead and high scalability. 

In practice, the significance of a bit flip strongly depends on its location, i.e., which bit in the floating point representation is affected.
Classical \acrshort{abft} has been extended to take into account floating point effects in the fault detection (checksums in finite precision) as well as in the fault correction and to recover from undetected errors (bit flips) in all positions without additional overhead \cite{moldaschl.etal:2017}.

\paragraph*{Iterative linear solvers}

\begin{figure}[htb]
    \centering
    \includegraphics[width=0.32\linewidth]{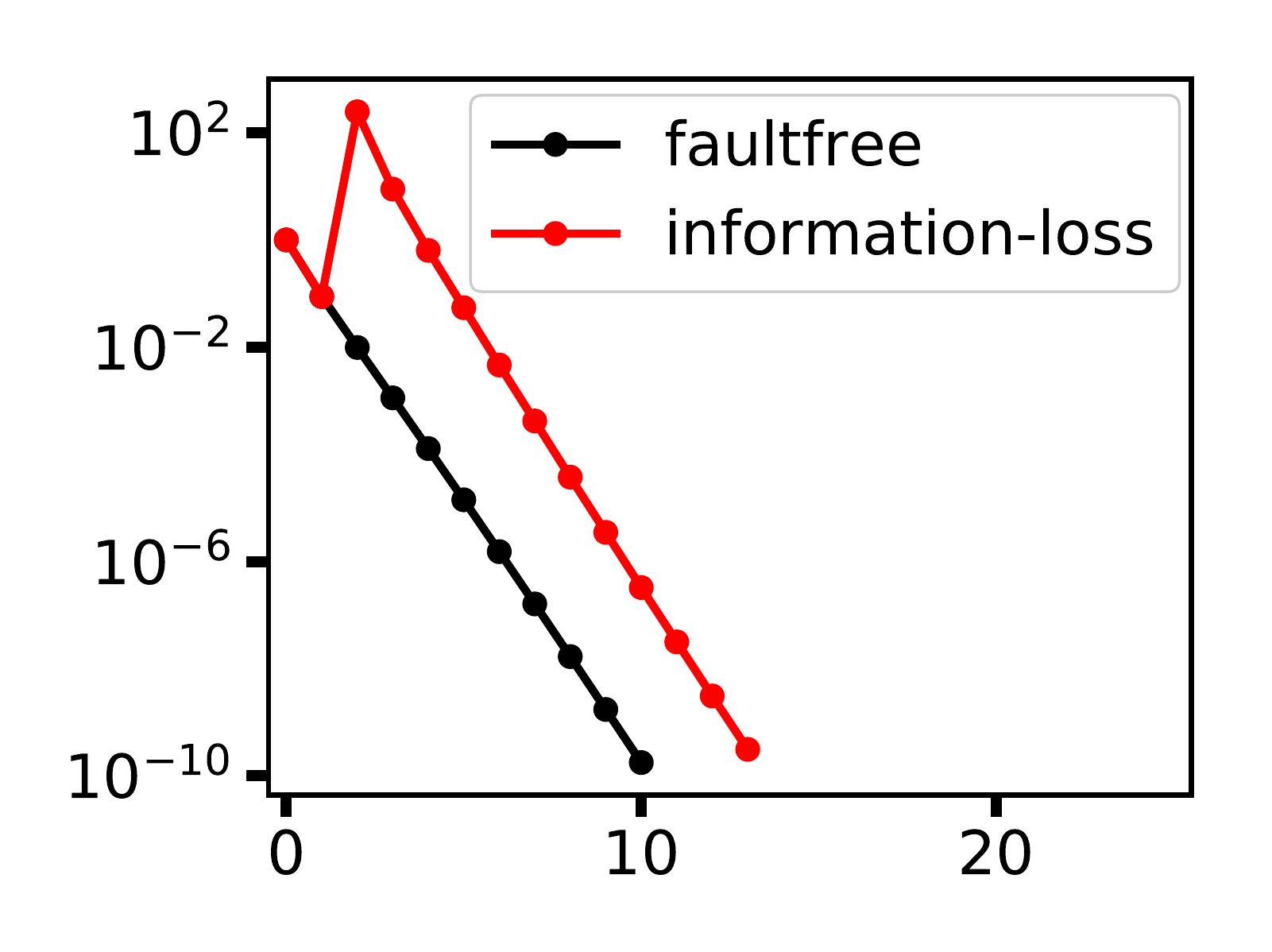}\hfill
    \includegraphics[width=0.32\linewidth]{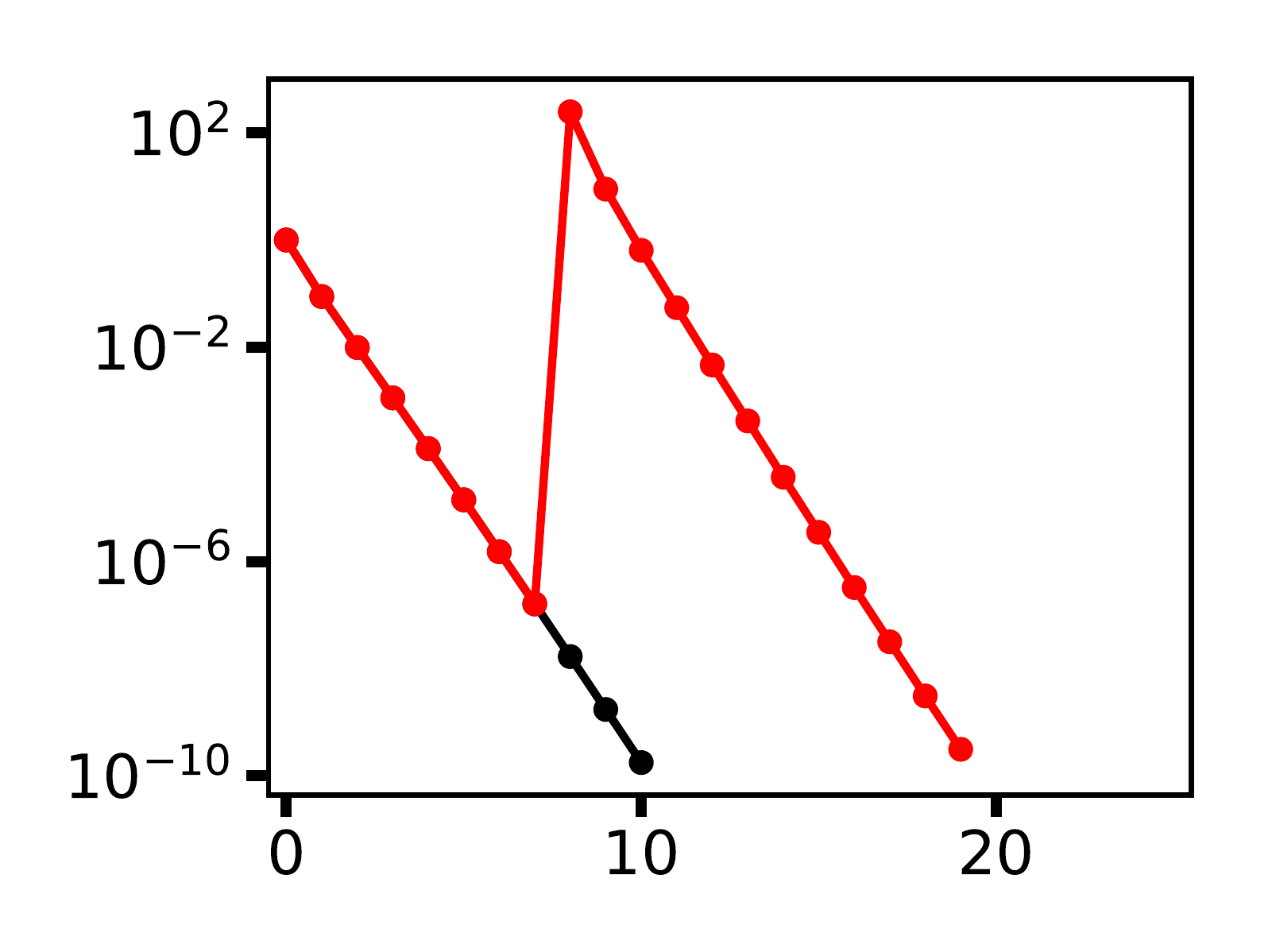}\hfill
    \includegraphics[width=0.32\linewidth]{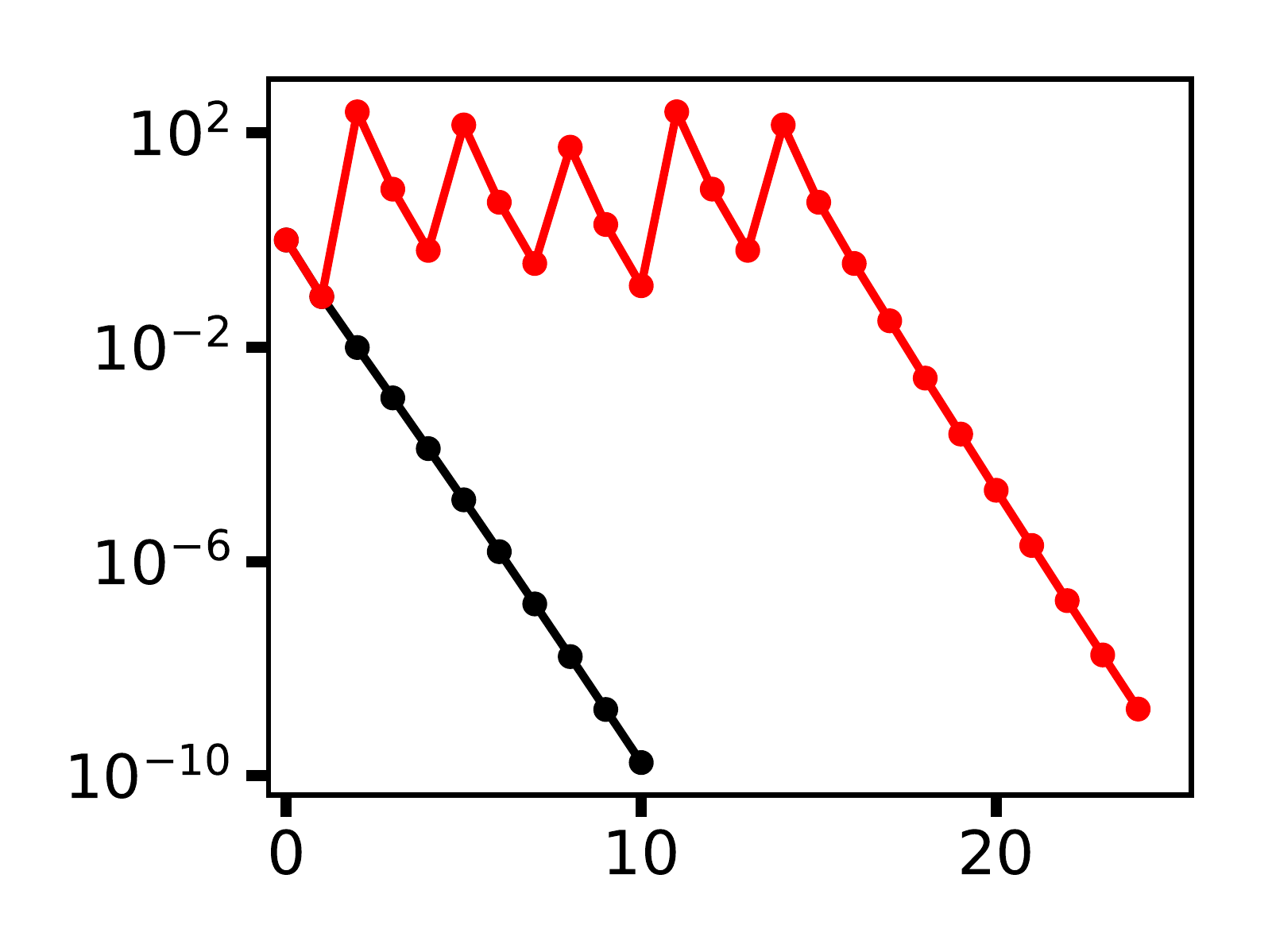}
    \caption{Convergence history of the residual norm as a function of the iteration count for three examples of information loss. From left to right: early, late, and multiple times}
    \label{fig:erroraware:typicalbehaviour}
\end{figure}

Iterative linear solvers based on fixed point iteration schemes are, in general, examples of error oblivious algorithms, as described in Section \ref{sec:error_oblivious}. 
The convergence history of the scaled residual norm observed within the iterative scheme often resembles the curves displayed in  Figure~\ref{fig:erroraware:typicalbehaviour}.
In this case the iterative scheme is a multigrid method, as in \cite{Goeddeke:2015:FTF, huber2016resilience}. 
The peaks in the residual occur after data has been lost 
and when the iterations are allowed to restart with
some form of replacement of the lost data.
In the simplest case, the lost data may just be re-initialized with
the value of zero,
and recovery techniques to obtain better solutions are discussed in
Section~\ref{sec:error_aware_case_studies}. 

It can be seen that, depending on when in the course of the iteration a small portion of the approximate solution suffers from an error,
we observe a delay in convergence, directly proportional to an increase in runtime. 
In the case where errors appear too often, the solver might not recover and other mitigation actions might have to be considered. 

Explicit recovery at the algorithmic level from undetected errors have 
been studied for iterative linear solvers~\cite{pachajoa2017resilience}.
In contrast to restarting, a number of algorithm based recovery strategies have been proposed, 
including approximate or heuristic interpolation methods~\cite{agullo:hal-01323192}.
An approach of exactly recovering the state of the iterative solver before the node failure has been investigated for the \acrfull{pcg} and related methods~\cite{pachajoa.etal:2018, levonyak.etal:2019}.
This also includes studying scenarios with multiple simultaneous node failures~\cite{pachajoa.etal:2019} and scenarios where no replacement nodes are available~\cite{pachajoa:2019}.

\newcommand{\ignore}[1]{}
\ignore{
\paragraph*{Combining checkpointing and numerical mitigation}

Shrink and substitute paradigms are tested to protect the GMRES Krylov solver against hard faults in \cite{ashraf:2018} and combined in a flexible approach featuring incomplete data recovery. An algorithmic rather than ULFM-based solution (see Section~\ref{sec:ULFM}) to recover from detected errors and shrinking processor pool is presented in \cite{pachajoa:2019}. Here, iterates and search directions of the conjugate gradient method are protected, and performance is achieved redistributing loads on reconstruction nodes rather than using replacement nodes. The related row permutations assume initial load balancing.

``Algorithm-based checkpointing'', i.e., hybrid forms between algorithm-based fault tolerance and in-memory checkpointing has been considered in \cite{pachajoa.etal:2019, levonyak.etal:2019} for variants of the \acrshort{pcg} algorithm. The basic idea is to identify the minimum amount of information which has to be checkpointed regularly in order to allow for the complete reconstruction of the correct state of the solver before the event of a node failure. After this reconstruction phase, which can be done very efficiently for several variants of \acrshort{pcg}, the solver can continue exactly as in the fault-free scenario.
}


\paragraph*{Approximated recovery and restart in sparse numerical linear algebra}

For matrix computations, eigensolvers or basic kernels such as iterative linear system solvers, 
some recovery ideas rely on forming a small dimensional linear algebra problem where the inputs are the still valid data and the unknowns are the lost/corrupted ones.
The outcome of this procedure is subsequently used to replace the lost/corrupted data and the numerical algorithm is somehow started again from that meaningful initial guess. 
The recovery procedure is tailored to the actual numerical algorithm. 
As an example, consider a fixed point iteration scheme for a linear system and suppose the lost data are entries of the iterate vector, the most dynamically evolving data in this computational scheme.   
Matrix entries of the iteration scheme related to the lost data, as well as some neighbouring entries, 
serve to build  the left-hand side of a linear problem (either a linear system or a least-square problem) while the right-hand side is built from valid data.
The solution of this small problem is then used to replace the corresponding lost entries of the iterate vector. 
The complete, updated vector is taken as a new initial guess when restarting the fixed point iteration.
If the data is not corrupted too often the classical convergence theory still applies and because the new initial guess incorporates updates from the calculations performed before the error was detected, the global convergence rate is not strongly affected.  
The method described in adaptive recovery techniques for extreme scale multigrid in  Section~\ref{sec:error_aware_case_studies} 
is an example application of this technique.
 
For numerical schemes based on nested subspace search,
such as Krylov subspace methods, closely related techniques have been successfully applied both for eigensolvers and linear solvers that further exploit the sparsity structure of the matrices to reduce the computational cost associated with the recovery procedure. 
At the cost of a light checkpoint performed once when starting the linear solver (mostly the matrix and the right-hand side vector in case of linear system solution) this mitigation approach has no overhead 
if the data is not corrupted during the solution computation.  
We refer to~\cite{Langou:2004:fault:recovery,agullo:hal-01323192,agullo:hal-01347793} for some illustrations on those numerical remedies in a parallel distributed memory framework and to~\cite{jcmA:15} where these ideas are exploited for a lower granularity of data loss in a task-based runtime system.
See Section~\ref{sec:infrastructure} for references relevant to task-based runtime systems.
 
 We also note that these ideas can be extended to hybrid iterative/direct numerical schemes, that have a domain decomposition flavor, where the recovery procedure can be enriched with additional features of the parallel numerical scheme such as redundancy or  properties of the preconditioners~\cite{agullo:hal-01256316}. They can also be extended to the time domain in the context of multilevel parallel-in-time integration techniques~\cite{speck2017toward}.
 

\subsubsection{Error aware algorithms for the solution of partial differential equations}\label{sec:error_aware_case_studies}

 The ideas introduced above in Section~\ref{sec:numalgos:erroraware:relatedwork} are application agnostic but naturally apply to linear systems arising from the discretization of a \acrshort{pde}. 
 In that latter case, more information from the underlying \acrshort{pde} can be closely tailored to intrinsic features of solvers such as multigrid. 
 In this section we discuss some research works on mitigation and containment that exploit the properties of \acrshort{pde}s to aid the recovery techniques. 
 We also present some mitigation processes that are only relevant in the \acrshort{pde} setting.
 


\paragraph*{Adaptive recovery techniques for extreme scale multigrid}

Some of the most efficient solvers of \acrshort{pde}, such as 
parallel geometric multigrid methods \cite{hulsemann2006parallel,gmeiner2015towards},
can be based on the exchange of ghost layers in a non-overlapping domain partitioning.
This automatically leads to a
redundancy in interface data 
between subdomains that 
in turn permits the design of an efficient two-step recovery strategy for iterative solvers. 
This is of particular interest in large-scale parallel 
computations. 
When each subdomain is large, then the ratio between the data on its surface and 
the volume data in its interior becomes small. 

When a processor fails, the information within one or several subdomains is lost.
For the recovery and continued solution,
the redundant ghost layer information is used
in a first step, to
recover locally either 
Dirichlet- or Neumann-type data for the subdomains. 
The global problem can then be formulated in two partitions,
the outer healthy subdomain
and the inner faulty subdomain, where the recovery must
reconstruct the lost data.
Both subproblems must be bi-directionally
coupled via the interface and the corresponding ghost layers of unknowns. 

After re-initialization, the corrupted and reinitialized data could
pollute the solution globally, meaning that the locally increased
error in the faulty domain can spread 
globally and thus also affect the healthy subdomain.
In order to avoid this pollution,
the communication between the healthy and faulty sub-problems is interrupted during the second step of the recovery process.
In the second step, we continue with the original iterative solver restricted to the healthy sub-problem and select a suitable one for the faulty one.
After some number of asynchronous iteration steps both sub-problems are reconnected, see \cite{huber2016resilience}, and the global iterative solver strategy is resumed.
Note that the reconnecting step is mandatory for the convergence of the iterative solver. 
The tearing step separating the subdomains
is mandatory to preserve the accuracy of the dynamic data in the healthy sub-problem, and without this step the corrupted data from the faulty sub-domain pollutes the global solution. 
Of critical importance for the performance of the method are the accuracy of the faulty sub-problem solver at re-connection time and the time spent in the recovery mode.
In the faulty domain, the lost data can be initialized with 0, or, alternatively,
compressed checkpointed data can be used as described in the following section on
compression techniques for checkpoint-restart.
Note, however, that with straight-forward compression techniques, compressed checkpoint data will only be useful
to recover the low frequency components in the faulty domain.
If the local recovery is performed with multigrid, then the low frequencies are
in any case cheap to recover, as compared to the cost of recomputing the lost
high frequency components. 

\begin{figure}[tb]
    \centering
    \begin{minipage}{0.3\textwidth}
    \includegraphics[width=\textwidth]{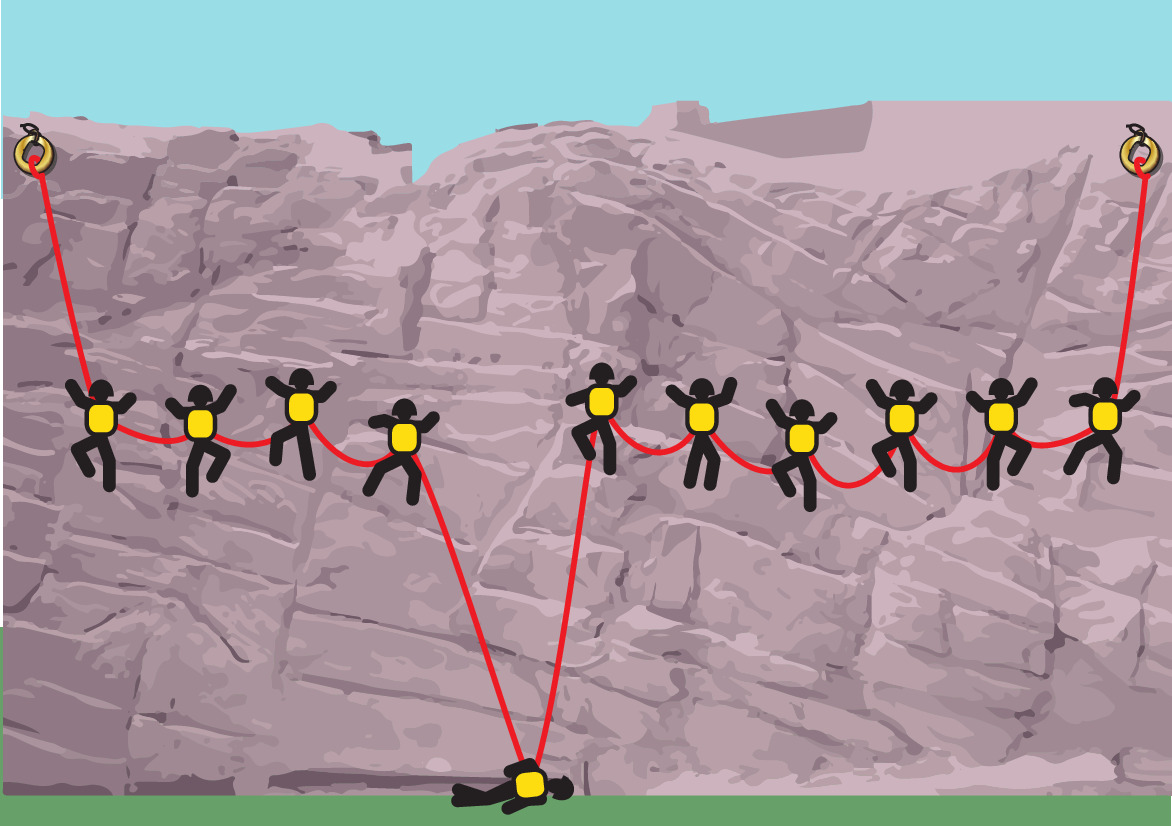}
    \end{minipage}
    \hfill
    \begin{minipage}{0.3\textwidth}
    \includegraphics[width=\textwidth]{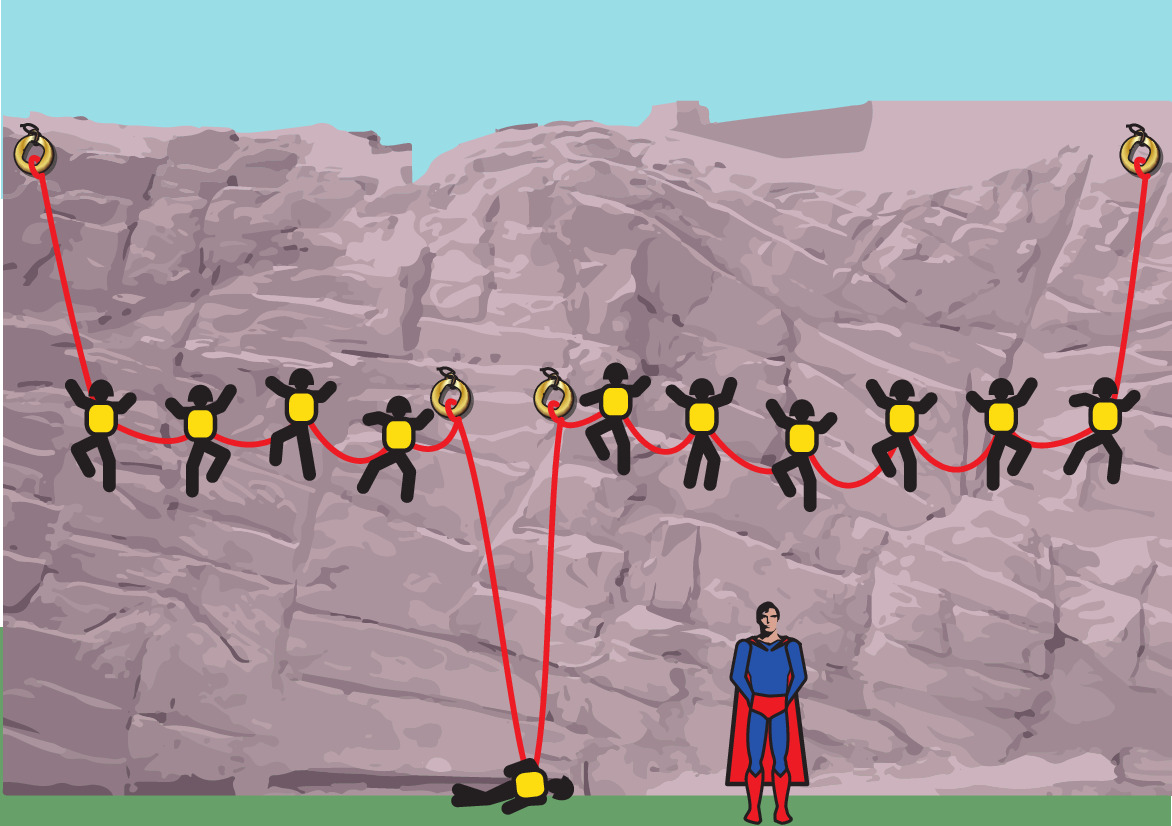}
    \end{minipage}
    \hfill
    \begin{minipage}{0.3\textwidth}
    \includegraphics[width=\textwidth]{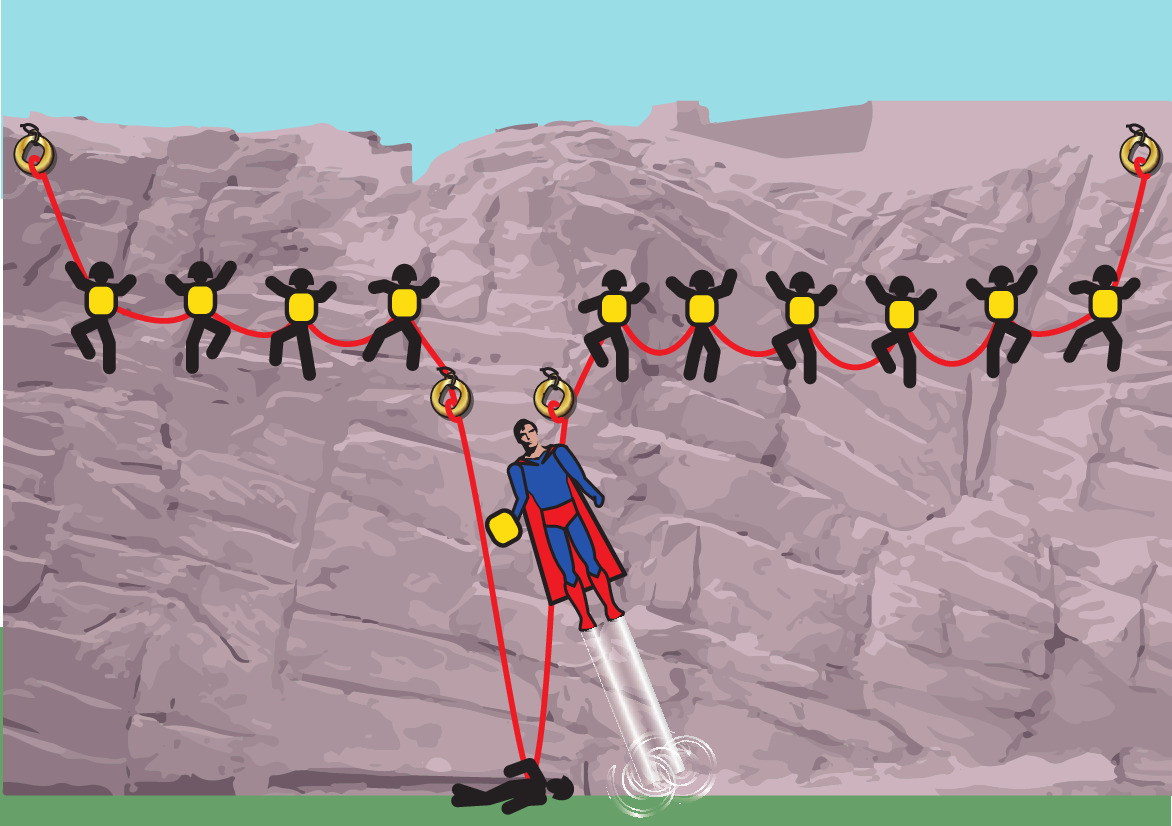}
    \end{minipage}
    \caption{Illustration of the steps in the adaptive recovery technique for extreme scale multigrid. 
    Left: A detectable error occurred. 
    Middle: The communication between the healthy and faulty sub-domains is interrupted. 
    Right: The original iterative solver restricted to the healthy domain continues while another suitable solver is asynchronously used in the faulty domain. 
    Once the solution in the faulty domain reaches a certain accuracy, the communication between the domains is re-enabled.}
    \label{fig:superman}
\end{figure}

The accuracy within a multigrid strategy can be easily controlled by a hierarchical sum of weighted residuals \cite{rude1994error}.
The overhead cost for the a-posterior error indicator is quite small compared to the overall solver cost. Having an estimate for the algebraic error in both sub-problems at hand, the re-connection step is determined automatically.
To speed up the time which is spent in the recovery, a so-called \lq superman strategy\rq\ is applied \cite{huber2016resilience}, see also Figure
\ref{fig:superman} for an illustration.
More resources compared to the situation before the fault are allocated to the faulty sub-problem. 
A short recovery phase in combination with carefully selected re-coupling criteria then guarantees a highly efficient fault-tolerant solver.

Of special interest is a massively parallel multigrid method as base solver.
In combination with the tearing and intersection approach for the recovery, it results in a hybrid approach. 
In case of a Stokes-type system, yielding after discretization a saddle point problem, the strategy can either be applied on the positive definite Schur complement for the pressure or, as it was done in \cite{huber2019adaptive}, on the indefinite velocity-pressure system. 
In that case an all-at-once multigrid method with an Uzawa-type smoother acting on both solution components turns out to be most efficient, see \cite{drzisga2018block}.
Numerical and algorithmic studies including multiple faults and large-scale problems with more than $5 \cdot 10^{11}$ degrees of freedom and more than $245 000$ cores 
have been demonstrated \cite{huber2016resilience,huber2019adaptive}.
The automatic re-coupling strategy is found to be robust with respect to the fault location and size and also handling multiple fault. 
In many scenarios a complete recovery can be achieved
with almost no increase in runtime and  while maintaining
excellent parallel efficiency.  


\paragraph*{Adaptive mesh refinement, load balancing, and application level checkpointing}

\acrfull{amr} functionality and load balancing require similar data linearization- and transfer functionality as is needed for application level checkpointing. 
This is exploited in the waLBerla framework \cite{schornbaum2016massively,schornbaum2018extreme,bauer2020walberla}
that features an object oriented design for composing coupled multiphysics simulation software.
waLBerla's load balancing is based on routines to transfer
simulation data between processors
so that functionality to serialize, pack, send, and unpack all relevant data is 
already available as a by-product of the \acrshort{amr} 
functionality. 
Note that the waLBerla software architecture imposes this structure for Eulerian mesh based data
as well as for Lagrangian particle-based models and it canonically  
extends to coupled Eulerian-Lagrangian multiphysics models.
For this to work transparently, the routines for migrating simulation data 
must be suitably encapsulated. 
Then this functionality can be used to write user level checkpoints
either on disk or in memory.
Note that writing checkpoints 
will inevitably imply overheads in memory consumption and communication time,
but that restoring a checkpoint is cheap, since it initially only requires re-activating
the redundantly stored data.
This is especially true when in-memory checkpointing is used as explored and analyzed in~\cite{kohl2019scalable}.
The simple restoration of checkpointed data may of course lead to load imbalance,
but the functionality to redistribute load is also available as part of the
parallel \acrshort{amr} functionality.
In this sense, user-level checkpointing can be based in a natural,
efficient, and straightforward way on the functionality of parallel
\acrshort{amr} algorithms combined with load balancing functionality.


\paragraph*{Compression techniques to accelerate checkpoint-restart for Kryloy-MG solvers}

Compressed checkpointing is a possibility to improve the efficiency of classical checkpoint-restart schemes, both in terms of the overhead to generate the checkpoints and to recover the data if an error occurs. The added efficiency mainly comes from a reduced memory footprint which is beneficial for communication and storage.
It is particularly efficient if the compression method is tailored to the target application. 
As an example, in-memory compressed checkpoints combined with \acrshort{lflr} (see Section~\ref{sec:numalgos:erroraware:relatedwork}) for iterative linear solvers, e.g., multigrid preconditioners in Krylov schemes, are described below. 

\emph{Lossy Compression:} As already mentioned in Section~\ref{sec:error_aware_case_studies}, paragraph \lq Approximated recovery and restart\rq, initially only the dynamical data, i.e., the approximate solution, are protected. 
Lossy compression allows a balance between the accuracy 
of the discretization error of the assembled system and the numerical error within the solver. 
Specifically in~\cite{Altenbernd:2020:hpcse}, the SZ library~\cite{di:2016,tao:2017,liang:2018} is employed, which prefers, by construction, structured data ideally associated with a structured grid.
Another important feature is that the compression accuracy can be prescribed and adapted to the situation. 
Unfortunately, a higher compression accuracy usually leads to a lower compression rate and higher compression time, which is crucial in terms of resilience overhead.

Note that multigrid can be interpreted as a lossy compression technique in itself, with a number of mathematical peculiarities that need consideration~\cite{Goeddeke:2015:FTF}.
Multigrid algorithms use a hierarchy of grids to solve linear systems in an asymptotically optimal way. 
This hierarchy can be used to restrict, i.e., lossily interpolate, the iterate from fine to coarse grids. 
Such a lower-resolution representation of the iterate can then be stored as a compressed checkpoint. 
Conversely, the multigrid prolongation (coarse-to-fine grid interpolation) operator is used to decompress the data. With only small additional computations, the multigrid hierarchy can also be used for error detection.

\emph{Recovery:} Several recovery techniques can be devised~\cite{Altenbernd:2020:hpcse}. 
As a baseline approach checkpoint-restart is mimicked and the global iterate is simply replaced with its decompressed representation, independently of the compression strategy. 
The second proposed approach follows the \acrshort{lflr} strategy and re-initializes only the local data that is lost on faulty computing nodes by using checkpoint data stored on neighbouring computing nodes. 
Contrary to the first approach, this is mostly local and only needs minimal communication to receive a remotely stored backup.
In particular, the recovery procedure itself does not involve the participation of other processes except those sending the checkpointed data. 
As a worst-case fallback when the backup data is not sufficient, a third  recovery approach is established, which is still mostly local. Here, an auxiliary problem is solved iteratively with boundary data from the neighbouring computing nodes.
This is similar to the adaptive 
recovery 
techniques for extreme scale multigrid from above or the
approximated recovery and restart of Section~\ref{sec:numalgos:erroraware:relatedwork}.\todUR{It is a bit difficult to ref the individual sections here since they have no numbering. Should we
add proper subsection structure? \textbf{DG: At the beginning of \lq our\rq\ paragraph, I made a slightly different suggesion how to handle this problem.}}
An auxiliary problem is constructed, either by domain decomposition overlap or the operator structure, and solved with an initial guess based on the checkpoint data to accelerate the iterative recovery phase. 
Experiments show that this approach can almost always restore the convergence speed of the fault-free scenario independently of the used backup technique, only the number of additional local recovery iterations varies. 
For more details, we refer to \cite{Altenbernd:2020:hpcse}.


\paragraph*{Resilience with sparse grids}

Resilience can be added on various abstraction levels of the algorithm. 
For PDE problems one traditionally adds resilience on the level of linear algebra operations, on the solver level for linear/non-linear equations, or on the time-stepping algorithm. 
However, this may in some cases not be coarse-grained enough to minimize the overhead of resilience techniques, especially when errors occur rarely. 
In \cite{obersteiner2017highly,heene2016massively, heene2018exahd, parra2015towards, parra16SDC, harding2015fault} the authors demonstrate a fault-tolerant framework for solving high-dimensional PDEs that applies fault tolerance on top of the individual PDE solver. 
The framework boosts the scalability of black-box PDE solvers while making it simultaneously resilient to faults by applying the sparse grid combination technique.
In this technique the PDE simulation is distributed over many coarse grids, which can be processed in parallel.
At regular intervals the results of these grids are combined 
to obtain the final sparse grid result.
In presence of faults the affected grids can be neglected and an alternative combination scheme is calculated via an optimization routine.
If too many grids are lost, the last combination result serves as an in-memory checkpoint to recompute the required grids. 
In \cite{obersteiner2017highly} it is shown that this lossy recovery provides very good results even with high error frequencies.
At the same time the parallel efficiency is only slightly affected. 


\paragraph*{Adaptive mesh refinement}

Adaptive refinement techniques in combination with finite element methods are well established for fault-free computations.
In terms of fault tolerance, this means that in addition to the assembled linear system, the geometric mesh structure must be protected.
This requires the reconstruction of the data structures containing the mesh hierarchy.
For the use of multigrid or multilevel methods, we also need to recover multiple levels of adaptive grid refinement after a fault has occurred.
The recovery process must take into account the intra-grid as well as the inter-grid data dependencies.

We refer to~\cite{stals2019} for a parallel adaptive multigrid method that uses a sophisticated dynamic data structures to store a nested sequence of meshes and the iterative evolving solution.
Stals demonstrates that it is possible to implement a fault recovery procedure that builds on the original parallel adaptive multigrid refinement algorithm~\cite{stals1995} in the case of a fail-stop fault.
It is assumed that a copy of the coarsest grid can always be accessed after a fault has occurred, i.e., it is stored off the processor.
The challenge in recovering an adaptively refined grid is that the mesh distribution changes during any load balancing procedures, i.e., the local information that was available during the original refinement process will have been modified or removed.
Nevertheless it is demonstrated that the neighbouring healthy processors contain enough intact information so that the necessary structure can be recovered to pass into the refinement routine.
In the case of uniform refinement, the original multilevel grid is recovered.
In the case of an adaptively refined grid, enough of the structure is recovered to re-establish the correct communication pattern allowing the solution process to run to completion, but potentially with reduced accuracy.
The neighbouring healthy processors will only contain enough information to guide the refinement around the edge of the recovered subgrid. 
Further refinement within the interior of the recovered subgrid may be required to improve the accuracy of the solution. 

These techniques were implemented with minimal disruption to the original code.
An example of one the few necessary modifications is that in the original code, communication was used to ensure that the elements were refined in the appropriate order to avoid degenerate grids.
In the resilient version of the the code that communication had to be removed as the refinement was restricted to the faulty processor. 


\subsection{Error oblivious algorithms}\label{sec:error_oblivious}



In this section, we give examples of algorithms that are error oblivious in the sense that they can recover without assistance from errors that do not occur too frequently.
For example, many fixed point iterative solvers are able to execute to completion if, e.g., a bit flip error occurs in the solution vector. 
However, every error likely increases the execution time of the algorithm. 
We thus define two quality criteria for error oblivious algorithms and use to assess the examples in the remainder of this section: (i) correctness, and (ii) efficiency in terms of execution time. 

Finding an algorithm that fulfills (i) and can also compete against error aware algorithms as described in Section~\ref{sec:error_aware_algorithms} remains an open problem. 

Error oblivious usually means that an error slowly  \lq leaves the system\rq\ during several iterative sweeps over the data. Error mitigation in error aware algorithms, on the other hand, requires specific measures to correct the error, and can only be applied when the error has been detected on a hardware, middleware or algorithmic layer, but removes the disturbance of the calculation process by the error immediately. 

We do not expect the error oblivious algorithms to be impervious to all types of errors. An iterative method may be not error oblivious if the error changed the matrix entries. This concept is defined as selective reliability, see Section~\ref{sec:numalgos:erroraware:relatedwork}.


\subsubsection{Gossip based methods}

A potentially interesting alternative in large-scale parallel environments that does not require any explicit error detection mechanisms utilizes gossip-based methods and their inherent resilience properties. 
Such algorithms by nature build up redundancy in the system and can thus can efficiently recover automatically from various types of faults/errors without any need to explicitly detect them. In particular, Gansterer et al. have studied and extended the resilience of gossip-based aggregation and reduction methods \cite{casas.etal:2019, niederbrucker.gansterer:2013, gansterer.etal:2011}. Based on these building blocks, they have developed and analyzed several more complex resilient numerical algorithms, such as orthogonalization methods~\cite{gansterer.etal:2013, gansterer.etal:2011}, eigensolvers~\cite{strakova.etal:2013}, and least squares solvers~\cite{prikopa.etal:2020}.

While the strong resilience properties and execution-time robustness of these approaches are promising, there is a certain price in terms of basic runtime compared to classical deterministic numerical high performance algorithms. It remains to be investigated whether they can be competitive in a fault-prone, but otherwise classical system with global view and centralized control. Their competitiveness can be expected to increase significantly if some of these classical properties have to be weakened at the extreme scale. 


\subsubsection{Fixed-point methods}
We view fixed-point methods as methods that converge globally when certain conditions are satisfied. 
For example, the Jacobi iterative schemes will converge for any initial guess if the matrix is diagonally dominant. 
Fixed-point based iterative methods are by design resilient to bit flips. However, the convergence delay can be significant. Anzt et al.~\cite{10.1145/2832080.2832081,ANZT2019100583} propose techniques improving the cost-robustness with little overhead.

A class of numerical algorithms that by design have properties attractive for resilience are asynchronous iterative methods~\cite{Baudet1978,spiteri1986,Bertsekas1983,bertsi1989,ebsmg1996,Szyld.98b,Frommer.Szyld.00,spiteri2020}.
In order to avoid misunderstandings, we point out that
this class of methods 
is unrelated to the idea of asynchronous dynamic load balancing~\cite{doi:10.1002/nme.6237}
as addressed in Section~\ref{sec:runtime}. 
Instead, asynchronous iterative methods, stemming from the concept of
chaotic iterations~\cite{chazan}, 
are fixed-point methods that seek the solution of a problem by independently updating subdomains -- which can be subdomains in the geometric sense, subsets, or individual components of the solution approximation -- according to some fixed-point linear or nonlinear iterative scheme.
A particularity of the asynchronous methods is that the independent updates neither adhere to a specific update order, nor synchronize in terms of a handshake with other updates, but still converge globally in the asymptotic sense.
In particular, these methods are robust with respect to some subdomains being updated at a much lower pace as each update just uses the most recent non-local information available. In that sense, asynchronous solvers can have good performance in unreliable environments where messages can be dropped or processes can become unresponsive for limited time.
Also, in cases where messages are corrupted (and corruption can be detected),
an asynchronous solver can simply drop such a message.
In cases where processes remain unresponsive, a mechanism is still needed to recover that process and its state, but the remaining processes can continue computing unchanged.
Therefore, asynchronous methods are 
somehow error oblivious.
%
%

With the increasing cost of global synchronizations, and the attractive properties concerning fine-grained parallelization and resilience against communication delays and errors, asynchronous methods have gained attention in particular for numerical computations~\cite{Szyld.01}. 
Chow et al.~\cite{DBLP:journals/siamsc/ChowP15,DBLP:conf/supercomputer/ChowAD15} developed an asynchronous algorithm for generating incomplete factorizations, Coleman et al.~\cite{DBLP:conf/springsim/ColemanSC17} further improved this algorithm by employing measures that reduce the runtime overhead when encountering errors.
More general is the idea of asynchronously updating subdomains in Schwarz decompositions.
In particular asynchronous restricted additive Schwarz methods and asynchronous optimized Schwarz methods have been identified to combine algorithm-inherent resilience with scalability on pre-exascale hardware architectures~\cite{DBLP:journals/pc/YamazakiCBD19,Magoules.Szyld.Venet.17,Glusa.etal.19,ElHaddad.Garay.Magoules.Szyld.20,Garay.Magoules.Szyld.18.3}.

Independently, asynchronous multilevel methods have been proposed and analyzed
under the name Fully Adaptive Multigrid method \cite{rude1993fully}.
Here the multigrid smoothing process is executed asynchronously so that it
can be employed for concurrent operations on different levels of the mesh hierarchy.
The iteration is executed in a Southwell style \cite{Southwell1949}
and is controlled by efficient hierarchical error estimators \cite{rude1994error}.
The parallel implementation \cite{rude1993mathematical} will
automatically correct errors.
More recently, asynchronous methods have been proposed for nonlinear multi-splitting~\cite{Szyld.Xu.00} and eigenvalue computations like Google's Pagerank algorithm~\cite{Kollias.Gallopoulos.Szyld.06}.
More recently, also the idea of asynchronously solving coarse-grid error correction equations has been investigated, leading to an asynchronous multigrid algorithm~\cite{WolfsonPou2019AsynchronousMM}.
While case studies reveal attractive properties, these newly developed asynchronous iterative methods (such as asynchronous multigrid) are not fixed-point iterations, and developing a convergence theory for those algorithms remains a challenge.

\subsubsection{Krylov subspace solvers}

A comprehensive overview about the use of selective reliability with Krylov methods in the presence of bit flips is  given in James Elliott's PhD thesis~\cite{Elliott2015ResilientIL}. 
Elliott  evaluates the CG and GMRES solvers with the algebraic multigrid preconditioner,
see also~\cite{Altenbernd:2020:hpcse} for a more recent study.
Coleman et al.~\cite{DBLP:conf/springsim/ColemanSC17} consider Krylov subspace solvers in combination with the incomplete ILU algorithm ParILUT. 
In~\cite{6877347} Elliot et al.\ investigate the effect of bit flips on the convergence of GMRES and propose strategies for minimizing the numerical impact.

The authors of \cite{BenacchioEtAl2020} present a monotonicity-based fault detection and correction procedure for a Generalized Conjugate Gradient Krylov solve and perform tests with manual fault injection. While the solver manages to converge even with large amounts of corrupted data, the basic recovery procedure speeds up convergence with minimal detection and correction overhead.

In \cite{Sao2013} the authors use a slightly different terminology and call their method numerically self-stabilizing, a term which originates in the context of distributed systems~\cite{dijkstra1974self}.
They introduce two error oblivious~\cite{dijkstra1974self} iterative linear solvers: one for the steepest descent and one for conjugate gradient. In the latter case, they consider necessary conditions for conjugate gradient to converge. Those conditions are borrowed from non-linear conjugate gradient~\cite{zoutendijk1970nonlinear} and are maintained in a correction step (typically performed every other ten iterations). The correction step does not explicitly correct errors, but re-computes quantities such as the residual at regular intervals. Therefore, we classify these methods as error oblivious instead of error aware. 



\subsubsection{Domain decomposition}

In~\cite{griebel2020stochastic} Griebel and Oswald use probabilistic analysis to model the effect of errors on the convergence of the classical overlapping Schwarz algorithm. 
They conclude that this method does indeed converge in the presence of errors.
Glusa et al.~\cite{DBLP:journals/corr/abs-1808-08172} mention that asynchronous domain decomposition methods are by definition fault-tolerant.
In \cite{Rizzi2016ftxs, RIZZI2018parcomp, Morris2016scala}, the authors discuss resiliency of a task-based domain decomposition preconditioner for elliptic PDEs. 
By leveraging the domain decomposition approach, the problem is reformulated as a sampling problem, followed by a regression-based solution update. The regression is formulated and implemented such that it is oblivious to corrupted samples. 
The authors combine this algorithmic approach with a server-client implementation based on ULFM, see Section~\ref{sec:ULFM}. 
They show promising results of this approach in terms of resiliency to missing tasks, corrupted data and hardware failure.


\subsubsection{Time stepping}
In~\cite{grout2017achieving}, iterative time-stepping using spectral deferred corrections are shown to be error oblivious at the cost of more iterations for the affected time-step. With error-estimators in place, time-integration techniques like Runge-Kutta methods will repeat the calculation of a time-step with smaller step sizes, if errors in the solution vectors are relevant~\cite{chen2013comprehensive}. This type of algorithms is resilient against errors in the solution vector of the new time step. Repeating the new time step with a reduced time step size is not the optimal measure in case of an error where repeating the step with the same time step size would be more efficient, but it leads to correct results. 





\section{Future directions}
\label{sec:future-directions}


In the final section we focus on the future direction of resilient algorithms.  We highlight what changes need to be made to current infrastructures to support the goals proposed by algorithm and application developers. Furthermore, we list those algorithms that are likely to come to the forefront as resiliency plays a more important role in the cost-benefit analysis of extreme scale simulations. And we mention some numerical methods that are yet to be fully explored in the context of resilient algorithms.

\subsection{Systems in support of resilient algorithms}\label{sec:future-systems}


We propose that resiliency will only be obtained by a multilayered approach incorporating operating systems, file systems, communication, programming models, algorithms, applications and education. 
In terms of the layers covered by infrastructure, the goal is to increase systems and delivered performance while keeping the detectable errors in the upper algorithm based layers constant.
We refer the reader to the recently published report by Radojkovic et al.~\cite{EU_HPC} for an overview of the needs of the next generation \acrshort{hpc} systems. 


\subsubsection{Error correcting codes}
%

Poulos et al.~\cite{PoulosFTXS2018} propose hardware \acrshort{ecc} assistance that can pass error syndrome information through to an application and use this to fix detected errors.  When an \acrshort{ecc} hardware error occurs that results in a \acrfull{due}, the \acrshort{ecc} hardware generates a syndrome which is a byproduct of the error detection.  For many \acrshort{ecc} schemes, a syndrome that corresponds to a DUE can be used to generate a list of possible corrections, one of which is taken to be the original uncorrupted data.  In this work, the authors show that this set is relatively small, meaning that the set of potential values for an application to search for their correct answer (before corruption) is also small.  They also study the error value distribution and show that for certain classes of problems it can be easy to identify obviously wrong answers.  For the application studied in \cite{PoulosFTXS2018}, work was done to correct a hydrodynamics application using conservation laws and average of neighbor cells.  This work requires changes to the hardware error reporting techniques and modification to the operating system to determine which application observed the DUE and pass it to an interrupt handler.

%


\subsubsection{Improving checkpoint/restart}

Independent of any additions, changes or new developments in the algorithmic or the system area, checkpoint/restart will remain a necessary component for any system. For one, no other technique can provide the needed resilience against full system outages; further, checkpoint/restart is also needed for developers to deal with limited job execution times and possible migration between systems or debugging purposes at large scale.

\paragraph*{Improving classical checkpoint/restart for homogeneous systems}

Observing the necessity of checkpoint/restart makes it critical to further optimize, enhance and support efficient checkpoint/restart mechanisms---even on classical, homogeneous systems---and provide users with library based solutions for core checkpoint/restart functionality. In particular, the following avenues should be pursued to optimize checkpoint/restart.

\begin{itemize}
    \item Use additional algorithmic techniques to be able to reduce checkpoint frequency.
    \item Reduce data to be written to disk by eliminating redundancy and possibly compressing checkpoint information. 
    Note that suitable data compression will typically require user-level knowledge,
    suitable interfaces must be provided.
    \item Overlap/Offload checkpoint operations to allow for asynchronous checkpoint/restart operations.
    \item Integrate checkpoint/restart with novel programming approaches to minimize checkpointable state.
    \item Keep the restart requirements local to the neighbour nodes of the failed node.
    \item Localize checkpoint data to own or localized nodes.
    This could be supported by local non-volatile memory, 
    as targets for checkpoint data. 
    While this has the potential to reduce communication, as it avoids remote data transfers, it may require additional hardware support to retrieve data from non-functional nodes, e.g., by accessing data through fast JTAG-like interfaces.
    \item In memory checkpointing.
    \item Exploit user-level knowledge for serializing, packing, compressing data,
    see e.g.~how existing \acrshort{amr} functionality 
    \cite{kohl2019scalable} can be exploited for efficient checkpointing in Section.~\ref{sec:soa-correction-incremental}.
\end{itemize}

\paragraph*{Checkpoint/Restart for heterogeneous systems}

In addition to classical checkpoint/restart for homogeneous systems, node-local checkpoint/restart support for heterogeneous systems will help containing error and failure propagation. Such support may be provided transparently to the application by the underlying infrastructure, such as \acrshort{gpu} drivers or task-based environments, or exposed in the programming model, such as OpenMP Offload \cite{diaz2018evaluating}.


\subsubsection{Scheduler and resource management}

Support for resilience, especially at the workflow-level, has a direct impact on resource management in \acrshort{hpc} systems and hence requires new developments in this area as well.

\paragraph*{Node-level parallelism} With increasing node-level parallelism, the impact of \acrshort{os} noise (typically caused by unpredictable interrupts) becomes even more important. Therefore, dedicated node-level resources are needed to exclusively run the \acrshort{os} and minimize the impact of \acrshort{os} noise on the multi-threaded application running on the other cores.

\paragraph*{Adaptive system and application load balancing} The batch scheduler needs to adaptively balance the system load onto the available resources, via seamless application migration. While the application needs to adapt to the capabilities of the newly allocated resources, if different from the original allocation, without incurring performance penalties. The former has typically been implemented via checkpointing and process migration~\cite{malleability:2007}. The latter has typically been implemented for 
applications that can adjust their granularity, e.g. from finer to coarser, depending on resource availability either triggered by the application or the system~\cite{malleability-invasive:2015}. When exposing and expressing parallelism in applications, in addition to accounting for and matching the multiple levels of hardware parallelism (nodes, sockets, cores), the decomposition granularity needs to be flexible to support evolvability and malleability and allow for adaptive load balancing at the application and system levels.

\paragraph*{Adaptive resource management} The batch scheduler in conjunction with the distributed runtime system employed by the application (e.g., MPI, Charm++, HPX) needs to support resources errors/failures and recover them without terminating the applications in the process. This approach should work both with rigid and moldable applications as well as with evolving and malleable applications. 




\subsection{Programming models with inherent resiliency support}
\label{sec:future-pms}


Certain applications and algorithms may naturally be resilient against errors. This may lend them as natural candidates for asynchronous parallel execution (via asynchronous many-task programming). While this mitigates the challenges associated with bulk synchronous parallel execution, asynchronous parallel execution may influence, in the presence of silent errors, the convergence rate of the numerical algorithms and might lead to incorrect results. 

%
%
Programming model and runtime support for resilience can offer transparent handling of errors and failures or can assist the application in handling them. Consistent programming model support for resilience based on realistic error/failure models is needed to properly handle such events with low overhead. Higher-level abstractions for programming resilient applications are needed to help with error/failure handling complexities and to offer reuse of concepts and codes across applications.

\subsection{Future directions for the solution of partial differential equations}\label{sec:future-grid-based}


In this section, we focus on discretizations for linear and non-linear partial differential equations as well as solvers for the resulting discrete and sparse systems of equations. We introduce a list of algorithmic properties that we found are, or can be, contributing to the resilience of the algorithms described in Section~\ref{sec:algorithms}.
Table~\ref{tab:grid-based-properties} lists these properties and indicates where we found relevant examples of how they can foster resilience for either linear or non-linear solvers or for spatial or time discretization.
In the following subsections we describe these examples in more detail and highlight the several (mutually related) properties that could be of interest in the context of resilient algorithms.
 

\begin{table}
\centering
\caption{Properties of numerical algorithms fostering or helping resilience}\label{tab:grid-based-properties}
\begin{tabular}{l|| cc}
categories & solvers & discretization \\
\hline \hline
redundancy                  & $\times$ & $\times$ \\
replication                 & $\times$ &        \\ \hline
hierarchical methods        & $\times$ & $\times$ \\
mixed precision             & $\times$ & $\times$ \\ \hline
error control               & $\times$ & $\times$ \\ \hline
locality-emphasizing schemes &        & $\times$ \\ 
asynchronous methods              & $\times$ & $\times$ \\
embarassingly parallel      & $\times$ &        \\ \hline
stochastic > deterministic  &        & $\times$ \\ \hline
iterative vs direct solvers & $\times$ &        \\ \hline
matrix-free / low memory footprint  & $\times$ & $\times$ \\
\end{tabular}
\end{table}


\subsubsection{Redundancy and replication}
\label{sec:future-gb-redundancy}

A failure that is not fixed by the system (hardware and middleware) typically results in a loss or corruption of data. 
To tackle this problem, redundancy techniques can be used to detect and recover from data corruption and data loss. The performance of these algorithms is usually measured in the amount of memory and computational overhead they entail, the detection rate of errors, the rate of false-positives they achieve, and the accuracy of the recovery. Optimizing these performance indicators should be of main concern for future algorithm design. One existing class of algorithms that apply redundancy are multiresolutional techniques such as multigrid and the sparse grid combination technique described in Section \ref{sec:error_aware_algorithms}. They inherently add redundancy through the hierarchical structure. Sparse grid combination techniques calculate the same solution on different anisotropic grids. The coefficients of the combinations  of the components grids can be recalculated if one or more nodes are lost due to faults. This redundancy of the component grids allows the algorithm to obtain an alternative approximation of the solution. However, if a component grid is distributed on too many nodes, then the approximation will fail if a fault occurs on any one of those nodes. Another class of algorithms add redundancy through recomputation with different models and configurations such as in ensemble or multifidelity techniques. A more straight-forward approach is to directly add redundancy through replication of certain algorithmic paths, cf. the following subsection on recalculation techniques.  

Depending on the underlying architecture, replication can be a competitive option to increase detected and undetected error robustness. 
If computation speed significantly outpaces memory access and communication, each operation can be executed multiple times while the data is still accessible in the RAM. This can be used for redundancy-based sanity checks of low-level operations or even for checksum-like approaches.

Overlapping data in parallel algorithms can serve as a starting point for mitigation, albeit not for detection. In the case studies explored in Section \ref{sec:error_aware_case_studies}, these are applied to elliptic PDEs, though an extension to other models should be feasible. Furthermore by even increasing the ghost layer size and thereby adding extra redundancy, other reconstruction possibilities might become possible. This could already be taken into account  during the domain partitioning process.





\subsubsection{Hierarchy and mixed precision}
\label{sec:future-gb-hierarchy}

Hierarchical discretizations have proven to be advantageous in various respects. Related notions are multi-resolution or multi-level discretizations, but also (recursive) sub-structuring in the engineering nomenclature of the \acrfull{fem}.
Built into the hierarchy are problem-inherent information and structures that are well-suited for modern hierarchy-based solvers. 
In \acrshort{fem}, for example, hierarchical bases carry information about both location and frequency, which leads to a 
special built-in redundancy that can be 
exploited for error detection (see Section \ref{sec:error_detection}). 
Therefore, from a resilience perspective, hierarchy should be a core paradigm for discretization design. 
This applies irrespectively of whether the hierarchical bases are formulated in the spatial ($h$) or the order ($p$) sense.

From a solver perspective, multigrid methods for elliptic and parabolic \acrshort{pde} problems are relevant approaches towards resilient numerical algorithms. They inherently act on different granularities, representations, scales, and levels and can be used to quantify differences between these levels. 
For local recovery, local multigrid methods are highly efficient, especially
when they can be accelerated with the superman strategy \cite{huber2016resilience}.
Additionally the low-resolution duplicates can be used for some kind of approximate recovery or minimal rollback like re-application of the smoother on a specific level in a multigrid scheme. 
Detection of errors within multigrid is often possible due to algebraic relations or on the basis of hierarchical multi-grid-inherent error estimates \cite{rude1994error, huber2019adaptive,altenbernd2016fault}, which hold true inside such schemes. As stated in Section \ref{sec:future-gb-redundancy}, the inherent redundancy incorporated in these algorithms is also beneficial. 

Mixed-precision arithmetics are typically used within the numerical solver parts to speed-up computations. However, the discretization can enable the flexibility to store data at varying precision. Examples for this are hierarchical approaches such as hierarchical bases, where a function value is stored as a hierarchical surplus only. As another example, the usage of wavelets in multiresolutional analysis can serve. In both cases, contributions of higher levels typically require less accuracy, as only the most significant bits contribute to the overall point values.


\subsubsection{Error control} \label{sec:future-gb-errorcontrol}


For many numerical methods, a wide range of classical a priori and a posteriori error estimation techniques are available, see among many others \cite{bank1993posteriori, rude1994error, quarteroni2008numerical, ainsworth2011posteriori, karniadakis2013spectral, lambert:1991,hairer:2008},  which constitute the basis of many adaptive numerical algorithms. 

Adaptive time discretization methods are the state of the art for \acrshort{ode} solvers, while, for \acrshort{pde} solvers, spatial adaptivity techniques are also widely used. Local time step adaptation is feasible in the framework of so called local time stepping or multirate approaches, where different components of the system can have different time step sizes, see  \cite{gear:1984, savcenco:2007, seny2014efficient, fok:2015, sandu2019class, delpopolo:2019, bonaventura:2020a}, which are however still far from mainstream for most applications. For \acrshort{pde} solvers, local spatial adaptivity techniques are also very common\cite{bangerth2012algorithms,bangerth2013adaptive}, but their incorporation in operational applications is still a research topic, see e.g. \cite{piggott2009anisotropic, berger2011geoclaw, leveque2011tsunami, muller2013comparison, tumolo:2015} for developments concerning
oceanography and numerical weather forecasting.   

The error estimations on which all these methods rely on also constitute the basis of an error detection mechanism, since some undetected errors, like bit flips on significant floating point digits, will result in errors exceeding the allowed error tolerances. 
To some extent, these techniques are also examples of \acrshort{abft} or error oblivious approaches, since bit flips and other silent errors occurring during the computation of the solution at the next time step or on a refined mesh could be automatically corrected by the repeated computations triggered by the error threshold violation. Furthermore, silent errors in the data at the current time or mesh level could be identified by the failure of the time step or mesh refinement to correct the error. 
 
Combined with other \acrshort{abft} strategies, adaptive discretization strategies based on error estimators can be a powerful and so far rather underrated tool for protecting a simulation from undetected errors in the solution vectors. On the other hand, error estimators should not be used as a black box for resiliency purposes. Indeed, errors can lead to severe over-resolution or, potentially, even under-resolution in space or time and the error estimators themselves could be affected by undetected errors.

As seen in Section \ref{sec:tec_err_info}, some iterative solvers for the solution of linear systems have invariants, such as monotonicity for Krylov solvers. 
These properties can be put to good use in devising resilience strategies, for example activating an additional restart of the Arnoldi procedure as soon as an increase in the residual norm is observed.

The idea of interval arithmetic is to compute bounds of intervals that always contain the exact result \cite{Alefeld1983,Kulisch2002}. 
Probabilistic methods for rounding error estimation \cite{vignes2004DSA,parker2001Monte,Frechtling:2015,denis2016,Verrou:2018} require several executions of arithmetic operations with different perturbations or different rounding modes (for instance three executions for Discrete Stochastic Arithmetic \cite{Eberhart2015REC}). 
With both approaches, the comparison of several computed results enables one to control rounding errors (or detect and mitigate actually wrong results).


\subsubsection{Locality, asynchronicity and embarassingly parallelism}
\label{sec:future-gb-locality-async-embpar}

One important aspect of resilient algorithms is error confinement as global dependencies propagate errors to other processors and complicate recovery. Locality-emphasizing numerical algorithms achieve this by limiting dependencies to local areas or completely removing them. Consequently, error mitigation can be limited to a local subdomain. Typical examples for these schemes are domain decomposition, which splits the domain into several subareas, and classical discretization schemes such as finite elements, finite differences and finite volumes.

As mentioned in Section \ref{sec:numalgos:erroraware:relatedwork}, domain decomposition schemes such as additive Schwarz methods, or substructuring-inspired FETI \cite{Farhatmethodfiniteelement1991}
or also the fully adaptive multigrid method 
\cite{rude1993fully} are naturally asynchronous and resilient to message loss.
In this context, we use the term asynchronous primarily in the sense of reducing the time synchronicity in parallel computations -- from communication-avoiding schemes via a reduction of synchronization points up to vastly decoupled schemes. 
Using this inherent property, a failure in a subdomain would result in a message loss that does not hinder convergence in other subdomains, because a global wait for a message update and synchronization are not necessary. 
In addition, asynchronous methods may better adapt to heterogeneous processors and networks than their synchronous counterparts as it has been shown in the context of Grid computing~\cite{myBCC05c,Chau2014}.
Both the localized and asynchronous approaches, achieve their impact through a decoupling of computations. Going further in this direction leads to embarrassingly or nearly embarrassingly parallel algorithms. These represent a group of algorithms where it is relatively easy to decouple subproblems in time or space. The subproblems can therefore be calculated completely independently, and errors do not propagate to other subproblems. Examples of such methods are Monte Carlo simulations and computations with the sparse grid combination technique. Since it is expected that only a few tasks will encounter errors and the scheduling is automatically balancing the load, the overall execution time does not suffer too much. Future algorithmic design should therefore aim at increasing asynchronicity and locality to move towards embarrassingly parallel problems.


\subsubsection{Stochastic}
\label{sec:future-gb-stochastic}

    Stochastic methods can be superior to deterministic methods when it comes to resilience. Stochastic methods do not require the program to take a deterministic path, faulty parts can be neglected or exchanged easily by other results. A popular example are Monte Carlo methods where we sample randomly in the computation domain and can simply neglect failed samples. Ensemble methods are examples where different instances or models of a concrete problem setting are computed. Even if one of these computation fails, the ensemble computation can still return a -- maybe slightly less accurate -- result. Stochastic elements can therefore help the future algorithm design to reduce the dependencies on specific results of the computation. These methods, however, need to be evaluated not just by highlighting their resilience properties, but also taking into account the cost of a single run: if a single run is expensive to complete, simply discarding it might be impractical. 
    
    

\subsubsection{Iterative methods} \label{sec:future-gb-iterative}

Iterative solvers may be viewed as inherently more robust than direct solvers because they do not compute their solution using a pre-defined sequence of numerical operations as direct solvers typically do. Indeed, by their nature, they perform a sequence of operations to update and improve their current approximation. If an error is encountered during computation, the probability of deleting this error or at least its effect may be higher than in a direct solver. Especially fixed-point-based methods (domain decomposition, relaxation, \ldots) may be viewed as inherently resilient as they have the property to always converge to the correct solution independent of the initial state (global convergence). Some errors may induce a low influence on convergence speed and can thus be safely ignored. In other cases, a restart -- optionally with recovery techniques -- may be employed to ensure both resilience and efficiency in terms of runtime.


\subsubsection{Low memory footprint -- matrix-free}
\label{sec:future-gb-matrix-free}

The classical approach to represent linear operators as sparse matrices
produces large amounts of static data which has to be restored upon failure.
Checkpoint-restart approaches feature high memory cost, naturally multiples of the storage needed for the solution vector. Algorithmic alternatives to checkpoint-restart require possibly complicated or costly re-assembly.
Matrix-free methods do not represent the operators as static data in the first place. 
Therefore, large sparse matrix data structures do not have to be restored upon failure as they are computed on the fly anyway. 
Extreme-scale applications will benefit from matrix-free approaches due to  their low memory footprint, also in terms of runtime, (due to high memory access cost) and higher limits for the overall problem size \cite{bauer2017two,bauer2018stencil}.

In addition to saving memory and, therewith, reducing the risk of memory corruption, matrix-free methods can also be combined with automatic code generation \cite{Lengauer:2020:ExaStencils} in a stencil-based approach, i.e., for finite difference methods on uniform structured grids. 
In such cases, the matrix entries may be \lq hard wired\rq\ into code, such as 5-point stencils for Laplace's equation. 
Automatic code generation provides a means to increase resiliency in the code generator or domain specific language and, thus, facilitate resilience aware software development. 

For finite element methods, one can use local assembly kernels
\cite{bauer2019large}.
Here, the trade-off between computation and storage and, in the future, resilience is relevant in particular for higher order elements. 

\subsection{The final mile: towards a resilient ecosystem} \label{sec:future-ecosystem}


The future directions described above will provide critical enhancements towards providing resilient computation for numerical simulations. Alone, however, they are insufficient, as they must be embedded in the larger ecosystem and in the efforts to make that ecosystem support such novel resilience approaches. This requires another set of crucial developments.


\subsubsection{Tools to support resilience software development}

Developers will need the right tools to support their algorithmic efforts. These tools, as they exist today, are often designed without faults and errors in mind and, therefore, do not sufficiently support the development of resilient systems. In particular, we identified three areas in which enhanced tool support for resiliency is needed: a) introspection to help track errors and failures along with their root causes, b) validation through controlled fault scenarios to enable targeted testing of new error mitigation features, and c) transformation to transparently add error and failure checks into codes.

\paragraph*{Tools for introspection}

Introspection is critical to ensuring early error detection and the timely activation of correction and mitigation mechanisms throughout the various layers of the software ecosystem.

\textit{System Monitoring:} 
%
Knowing about the health state of a system requires monitoring it and understanding its behavior. Future work needs to focus on scalable system monitoring, real-time analyzes of system monitoring data, and autonomous decision making on corrective actions for self-aware resilient systems. In order to gain a deeper understanding, types of monitored data should be homogenized across system and sites, and, if possible, sanitized logs should be available to the community.


\textit{Application and Data Structures Monitoring:} Applications need to automatically monitor their performance and correctness with the use of tools. The tools can be developed in abstraction, at the compiler-level, or at the runtime-level. 

\paragraph*{Tools for validation}

Currently, there are no standard tools to test the correctness and performance of resilient algorithms under undetected errors and fault.  This is due to a lack of fault injection tools that reflect realistic situations.  DeBardeleben et~al.~\cite{FSEFI} have developed a hardware error simulator tool to understand the behavior of numerical algorithms under faulty hardware with a great accuracy, but this approach cannot evaluate the execution time of resilient algorithms at scale.  Vendors provide fault injection
tools~\cite{4709308,DBLP:conf/ispass/HariTSKE17} for better execution efficiency, compromising the accuracy of the hardware behavior.  Compiler approaches or other in-house error injections~\cite{calhoun14flipit,georgakoudis2017refine}  could allow the program to execute as efficiently as the original binary, but the correctness is further compromised.  There are also tools that can analyze an application's vulnerability very quickly but do not actually produce the application's faulty output.  One technique for this, DiSCvar\cite{discvar}, uses algorithmic differentiation and exposes how changes to each variable impact output results.  It is important to note that these techniques do not actually produce that corrupted output. Hence, they are very fast but they may not be useful to developers looking to explore precisely how corruption changes their application.  It is likely that a combination of these techniques, which identify most critical regions of an application coupled with fault injection at those locations, may serve as a good compromise between the two techniques.

Any novel approaches that fill the gap between the accuracy and execution efficiency of error injections will facilitate the code development of resilient algorithms, and the new tools should be built with the existing continuous integration infrastructure.  Such tools likely require hardware knowledge that is considered intellectual property  by the semiconductor vendors.  However, efforts which explore this space using open hardware technologies (RISC-V, Sparc, etc.) can shed light on this space but may be of varying usefulness when application developers look to understand how their applications will perform on hardware that has not been fault injected at the register transfer  or microcode level.

\paragraph*{Tools for code transformation}

 Compilers are able to generate binaries with resilience capability as suggested in the work by \cite{reis2005swift}; the generated binary instruments redundant computation, register allocations to enable error detection and correction during program execution.
 The recent work by Lin \cite{lin2017simd} leverages LLVM to generate SIMD instructions to perform redundant computation and verification.  Source-to-source code transformation has been proposed to enable triple modular redundancy in loops \cite{lidman2012rose} and automatic instrumentation of checkpointing \cite{rodriguez2010compiler}. Similarly, this idea can be extended to redundant threading for error mitigation, facilitated with OpenMP-like programming language extension \cite{hukerikar2014redundant}.  These approaches automatically introduce resilience with some performance penalty, preventing the users from selective adaptation of resilience for performance optimization, and these redundant computations are benefited from the memory hierarchy, preventing doubling (or tripling) of the execution time.

In addition to such specific systems that support the addition of resilience to existing codes, automated generation of code, e.g., via \acrfull{dsl} can help with the transparent support of resilient computation. Examples for this can be stencil generators, as already discussed in Section~\ref{sec:future-gb-matrix-free}.


\subsubsection{User/Programmer education}

According to the system log study by~\cite{di2014lessons}, many application job failures are triggered by the mistakes of the users such as script errors and program bugs including excessive file and thread creations.  This means that better software engineering practices and training of users should be pursued with similar efforts to the deployment of resilience strategies.

The Exascale Computing Project (ECP) by the US DOE has made a substantial investment on educating tools, software engineering and \acrshort{hpc} system usage for a variety of the users.  Additionally, the scientific and mathematical library teams in the ECP have introduced software engineering policies~\cite{xSDK} to improve the software quality, documentation and testing process for better interoperability and composability of multiple library packages.   This activity, though not directly relevant to resilience, will gradually help to reduce application errors and failures for large scale \acrshort{hpc} systems. 

\section{Conclusions}
\label{sec:conclusions}

This article presents a snapshot of current research on resilience for extreme scale computing.
It has grown out of the Dagstuhl seminar 20101 held March 1-6, 2020,
bringing experts from the field together on the topic
\emph{Resiliency in Numerical Algorithm Design for Extreme Scale Simulations}.
This seminar became a starting point to develop a synthesis between the system perspective on resilience and the algorithmic perspective.

While resilience is undoubtedly an issue for extreme scale computing, 
it is less clear what algorithms on the user or application level can contribute to mitigate faults.
The seminar provided ample room to discuss these topics and thus became
the starting point for this article.
Many diverse aspects were found to be relevant, that require a holistic and multidisciplinary approach involving different and complementary scientific communities. 

In particular, it clearly appeared that a fundamental distinction lies in whether faults are detected or not,
and if they are not automatically detected, whether they are detectable. 
If they are, algorithms can often be developed to detect errors
and in a second stage to correct them.
It was found that some algorithms are naturally tolerant against faults
or have the intrinsic feature to be error oblivious.
They can thus be naturally applied on a system subject to errors.

Besides redundancy and checkpointing as classical techniques to mitigate faults,
new algorithm-based resilience techniques have been developed 
for several classes of numerical algorithms. 
This includes linear algebra and solvers for partial differential equations,
two classes of algorithms that are prominent in many scientific
workloads on supercomputers.
Some of these mitigation methods show remarkable success in the sense that
faults can be compensated algorithmically by recovery procedures
with only little extra cost in time or in silicon.
On the other hand it also becomes clear that integrating such techniques
in a computational infrastructure is still facing many obstacles. 
This includes the still poorly defined interface between user-level 
fault mitigation techniques and system level functionality, 
as, it is, e.g., necessary to reliably and
quickly detect a device (core, memory, ...) failure on a large parallel machine.

Despite its breadth, the article is far from being comprehensive.
The selection of topics is a subjective overview of
current research in the field of resilience for extreme scale computing 
and it delivers an outlook into possible and promising future research topics and solutions.

\bibliographystyle{plainurl}


\end{document}